\documentclass[5p]{elsarticle}

\usepackage{lineno}
\usepackage{adjustbox}
\usepackage{graphicx}
\usepackage{amsmath,amssymb}
\usepackage{mathabx}
\usepackage{mathtools}
\usepackage{algorithmic}
\usepackage{algorithm}
\usepackage{xcolor,soul}
\usepackage{hyperref}
\usepackage{empheq}
\usepackage{csquotes}
\usepackage{tabularx}
\usepackage{float}
\usepackage[english]{babel}
\usepackage{booktabs}
\usepackage{enumerate}
\usepackage{lipsum}
\usepackage{url}
\usepackage{bm}
\usepackage{times}
\usepackage{comment}

\usepackage[firstpageonly=true]{draftwatermark}	

\modulolinenumbers[5]

\journal{Journal of Process Control}

\newcommand\scalemath[2]{\scalebox{#1}{\mbox{\ensuremath{\displaystyle #2}}}}

\newtheorem{proposition}{Proposition}
\newtheorem{remark}{Remark}
\newtheorem{definition}{Definition}

\newcommand{\sss}[1]{_{\scriptscriptstyle #1}}

 \newcommand{\rev}[1]{#1}
\usepackage{color}

\hypersetup{
	colorlinks,
	citecolor=black,
	filecolor=black,
	linkcolor=black,
	urlcolor=black
}

\begin{document}
	\begin{frontmatter}

		\title{On Recurrent Neural Networks for learning-based control:\\
		       recent results and ideas for future developments}
		\author[First]{Fabio Bonassi}
		\author[First]{Marcello Farina}
		\author[First]{Jing Xie}
		\author[First]{Riccardo Scattolini\corref{corr1}}

		\address[First]{Dipartimento di Elettronica, Informazione e Bioingegneria, Politecnico di Milano, via Ponzio 34/5, 20133 Milan, Italy (e-mail:~\{name.surname\}@polimi.it)}
		\cortext[corr1]{Corresponding author}
				
		\begin{abstract}
		This paper aims to discuss and analyze the potentialities of Recurrent Neural Networks (RNN) in control design applications.
		The main families of RNN are considered, namely Neural Nonlinear AutoRegressive eXogenous, Echo State Networks, Long Short Term Memory, and Gated Recurrent Units.
		The goal is twofold.
		{Firstly, to survey recent results concerning the training of RNN that enjoy Input-to-State Stability (ISS) and Incremental Input-to-State Stability ($\delta$ISS) guarantees.}
		Secondly, to discuss the issues that still hinder the widespread use of RNN for control, namely their robustness, verifiability, and interpretability.
		The former properties are related to the so-called generalization capabilities of the networks, i.e. their consistency with the underlying real plants, even in presence of unseen or perturbed input trajectories.
		The latter is instead related to the possibility of providing a clear formal connection between the RNN model and the plant.
		In this context, we illustrate how ISS and $\delta$ISS represent a significant step towards the robustness and verifiability of the RNN models, while the requirement of interpretability paves the way to the use of physics-based networks.
		 The design of model predictive controllers with RNN as plant's model is also briefly discussed.
		 {Lastly, some of the main topics of the paper are illustrated on a simulated chemical system.}
		\end{abstract}
		
	\end{frontmatter}

% d\setchangekey{bonassi2022recurrent}{104}

 \DraftwatermarkOptions{%
 angle=0,
 hpos=0.5\paperwidth,
 vpos=0.95\paperheight,
 fontsize=0.012\paperwidth,
 color={[gray]{0.2}},
 text={
   \parbox{0.99\textwidth}{\copyright \, 2022. This manuscript version is made available under the \href{https://creativecommons.org/licenses/by-nc-nd/4.0}{CC-BY-NC-ND 4.0 license}. This manuscript has been published at Elsevier Journal of Process Control. Published article available at DOI \href{https://doi.org/10.1016/j.jprocont.2022.04.011}{10.1016/j.jprocont.2022.04.011}.}},
 }

\section{Introduction} \label{introduction}
The increasing possibility to collect measurements from the plants, the availability of sophisticated machine learning techniques capable of extracting information from large data-sets, and the availability of openly available tools to effectively train and deploy these algorithms are redrawing the landscape of data-driven control design.
In recent years, starting from different paradigms, a multitude of learning-based data-driven control strategies have been proposed, such as Koopman operator-based system identification \cite{korda2018linear}, set membership identification \cite{terzi2019learning}, Bayesian identification \cite{PigaBemporad}, and Gaussian processes learning \cite{hewing2019cautious}, also with focus on the stability properties entailed by these approaches \cite{aswani2013provably, berberich2020data}.
The clear reason behind this interest lies in the many potential advantages of these approaches over traditional model-based algorithms that require knowledge of a physical model of the plant, including the possibility to reduce the time and cost associated with physical model's design, tuning, and adaptation to new plant's operating conditions. 
 
 Starting from the eighties of the last century, Neural Networks (NN) have gained an increasing popularity in the Control Systems community for the design of data-driven control algorithms, especially when the system under control is characterized by a nonlinear behavior, which prevents one from using standard linear model structures, such as ARMAX and OE models \cite{schoukens2019nonlinear, narendra1996neural, hunt1992neural}; see \cite{aggarwal2018neural} for an up-to-date introduction to NN. 
Over the years, the use of NN for control has been advocated by many theoretical contributions, see e.g. \cite{sontag1997complete,levin1993control, levin1996control}, and by many accounts of applications, see e.g. \cite{willis1992artificial, himmelblau2008accounts}.
The wide variety of approaches can be arranged in six categories, which are briefly outlined in the following. 
For a detailed overview, the interested reader is addressed to \cite{hou2013model}. 

\subsection*{NN as model of a plant} 
 The most common approach consists in using NN as models in black-box identification procedures.
That is, starting from the input/output data collected from the plant, a NN architecture is chosen, and its parameters, named weights, are tuned during the so-called training procedure, with the goal of finding a set of weights for which the NN constitutes an accurate model of the system.
This model is then used for the design of model-based control architectures, such as Model Predictive Control (MPC).
Examples of this strategy in the academia are \cite{hunt1992neural, wu2019machine, terzi2021learning, bonassi2021nonlinear}, while accounts of applications in the industry are \cite{nagy2007model, lanzetti2019recurrent, wong2018recurrent}.

\subsection*{NN as a part of a grey-box model} 
Physical models are often characterized by {first-principle} equations featuring terms that are unknown or difficult to model, typically functions of other internal variables.
NN can be used to effectively model such terms from experimental data, thus blending classic physical modeling with learning.
This approach, formalized in \cite{karpatne2017theory} under the name of Theory-Guided Data Science, allows to avoid complex and over-parametrized black-box models, and it typically achieves enhanced interpretability and generalizability outside the identification domain \cite{schoukens2019nonlinear}.
% On the one hand, this allows to avoid the derivation and use, in the models, of overly-complex functions, which are often difficult or even impossible to derive analytically. On the other hand, the advantage with respect to pure data-based modeling is better generalizability outside the domain of identification and interpretability. This approach is formalized in \cite{karpatne2017theory} under the name of Theory-Guided Data Science.
Examples of this strategy are \cite{pozzoli2020tustin} and \cite{cranmer2020lagrangian}, where the high-level knowledge of mechanical systems is complemented with the use of NN to learn part of the nonlinear state update functions. 
A similar approach is proposed in \cite{hosen2011control}, where NN are used to learn the kinetic parameters of a CSTR system.

\subsection*{NN as model of the uncertainty}\label{sec:NN:uncertainty}
In case a preliminary model of the plant is available, be it identified from data or derived from physical equations, a NN can be used to model its uncertainty \cite{wu2020process}.
This allows to refine the existing model and to improve its accuracy.
The uncertainty, modeled by the NN, can then be taken into account by designing a control algorithm, such as robust MPC \cite{RawlingsBook}, that guarantees robust stability properties.
This strategy is most commonly applied in combination with linear models; in these cases, a robust controller is designed for the nominal linear system, while NN are used to learn the linearization error \cite{forgione2020model}.
Similarly, in \cite{zhao2019system} a nonlinear model of the system is derived from physical equations, and a NN is used to learn the residual modeling error. 
It should be noted that this approach allows, in some sense, to generalize to the nonlinear setting classic methods of unstructured \cite{Ljung2000} and structured  model error estimation, which can be performed for example with set-membership algorithms, \cite{Milanese98}.

\subsection*{NN as approximators of computationally-intensive control laws} \label{sec:NN:approximateMPC}
When a reliable model of the system is available, but the online computational burden of the designed control law is excessive, the interpolative power of NN can be leveraged to approximate such control law offline.
This approach is predominantly pursued when an MPC control law is adopted, which requires to solve a {potentially cumbersome} optimization problem at each time step.
While for the specific case of linear models, under mild assumptions, the MPC state-feedback law can be put in an explicit exact form \cite{alessio2009survey}, this task is generally relaxed for {nonlinear and/or nonlinearly constrained systems}, and an accurate approximation of the control law is sought.
To this purpose, NN are strong candidates, owing to their universal approximation capabilities and extremely low online computational cost, see e.g. \cite{parisini1995receding, cavagnari1999neural, karg2021approximate}.
Remarkably, it has been recently shown that these approximations can preserve the closed-loop stability \cite{hertneck2018learning} and fulfill input constraints \cite{kumar2021industrial}.
\rev{Similar approaches are those based on the Internal Model Control strategy, in which a NN is used to approximate the inverse of the model \cite{rivals2000nonlinear}, which enjoy a low computational burden while preserving the closed-loop stability \cite{bonassi2022imc}.}

\subsection*{NN as controllers directly synthesized from data}\label{sec:NN:direct}
A different approach is to directly derive the control law from the input/output data collected from the plant.
{This family of approaches enjoys a florid history for unconstrained linear systems, see e.g. \cite{terzi2019learning, campi2002virtual}; related strategies have been recently proposed for nonlinear systems \cite{tanaskovic2017data}.
In this context, learning NN controllers from plant's data has been investigated in \cite{yan2016data, radac2018data, DAmico2021}, as they not only lead to superior performances compared to linear control structures, but also, as shown in \cite{DAmico2021}, they can be designed to fulfill input constraints.}

\subsection*{NN for Reinforcement Learning}\label{S5}
Finally, NN have been often used in deep Reinforcement Learning (RL) strategies,  {see \cite{ozalp2020review, lewis2012reinforcement} for more details.
In the deep RL setting, a NN control law is learned to maximize a reward function throughout several closed-loop simulations.
To this end, a simulator of the system under control is generally needed, which usually limits the applicability of the approach. 
When a model of the system is available, a further possible domain of application of RL is to obtain an approximate solution to the nonlinear Hamilton Jacobi Bellman equation, whose exact solution can hardly be obtained, see e.g. \cite{kim2020model}.
}
% For instance, in \cite{lewis2012reinforcement} NN have been exploited to describe the cost function of the minimization problem associated with a RL scheme. 
% In general, nonlinear optimal control based on NN has been used to obtain the approximate solution to the nonlinear Hamilton Jacobi Bellman equation, whose exact solution can hardly be obtained. 
% \hltext{In these cases, adaptive NN dynamic programming can represent a valuable solution, see for instance \cite{bai2019adaptive}, where the proposed RL approach is based on the approximation of the utility function and of the unknown function by NN.} \footnote{FB: Non ho capito questa frase}
\bigskip

\begin{figure}[t]
\centering
\includegraphics[width=0.75\linewidth, clip, trim=0cm 1mm 0cm 0cm 0cm]{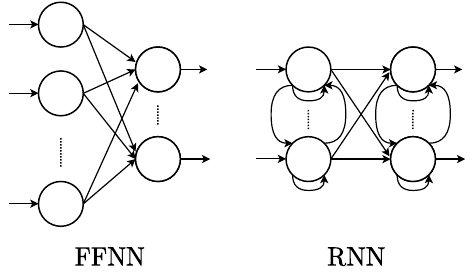}
\vspace{-2mm}
\caption{{Illustration of FFNN and RNN architectures. The latter are stateful network characterized by internal loops, which are absent in the former.}}
\label{fig:rnnscheme}
\end{figure}

In this paper, we focus on the use of neural networks as models of dynamical systems, learned from the plant's data and then used for model-based control design.
In general, {this strategy does} not merely boil down to a {model-based} control design problem, since the choice of the model's architecture and its identification are crucial ingredients to achieve satisfactory closed-loop performances, let alone the closed-loop stability.
Identifying suitably accurate, safe, and possibly interpretable models that represent an optimal compromise between simplicity and accuracy is hence the main challenge of {this approach}.
% \hltext{At the same time, an accurate characterization of the model inaccuracy should be provided in some terms (e.g., as structured or unstructured uncertainty, as stochastic or bounded additive disturbance, etc.), to be properly compensated or rejected by the controller.}\footnote{FB: Questa frase mi lascia perplesso, perché noi di questo problema non ci occupiamo.}

Of all the possible models and algorithms that can be used to identify a plant model, NN provide state-of-the-art performances and applicability to general classes of nonlinear systems \cite{schoukens2019nonlinear}.
The idea of harnessing the representational power of NN to identify dynamical systems  has a long and florid history, studded with a wide variety of architectures proposed for such task \cite{forgione2020model}.
A traditional approach consists to use Feed-Forward Neural Networks (FFNN) as one step-ahead predictors for the future output trajectories given past input and output data \cite{nagy2007model, piche2000nonlinear}, where FFNN are static and memoryless NN characterized by an unidirectional flow of information from the input layer to the output layer, through a sequence of hidden layers.
While this approach has been widely adopted in academia \cite{levin1993control, levin1995identification} and in industry \cite{himmelblau2008accounts, ali2015artificial}, there is nowadays consensus that these FFNN predictors lead to a poor accuracy in representing long-term dynamics, due to their inherent lack of memory.

For this reason, dynamical forms of NN, the so-called Recurrent Neural Networks (RNN), are typically adopted. 
The main characteristics of RNN is the presence of internal loops in the paths from inputs to outputs, {see e.g. Figure \ref{fig:rnnscheme}}, which correspond to the states of the dynamical system representing the plant under study.
A first simple RNN architecture is that of the so-called Neural Nonlinear ARX (NNARX)  \cite{bonassi2021nnarx,  sastry1994memory},  widely employed mainly thanks to their simple structure and training.
This architecture makes use of FFNN as regression functions embedded in a NARX setting, and it is trained minimizing the simulation error, thus obtaining a model able to mimic the system's free-run simulation.
The joint use of MPC and NNARX models, or FFNN predictors, has been considered in \cite{nagy2007model, hosen2011control, piche2000nonlinear}, with the aim of operating the system in a broader region compared to traditional linear models.

A second RNN architecture is that of the so-called Echo State Networks (ESN), originally introduced in \cite{jaeger2001echo}.
{The peculiarity of these networks is that they are trained by solving a least-square problem, for which efficient and provenly converging algorithms are nowadays available.}
ESN have already been used for control design in \cite{pan2011model, ploger2003echo}, while in \cite{armenio2019model} an MPC law with stability guarantees for ESN models has been devised.

The most advanced RNN architectures fall into the category of gated RNN, in which the internal loops are regulated by the so-called gates, that makes them significantly more suitable for learning dynamical systems.
In this group, the most popular architecture is that of Long Short Term Memory (LSTM) networks \cite{hochreiter1997long}, which have been recently considered in many control-related problems, see e.g. \cite{terzi2019learning, wu2019machine, terzi2021learning}.
Equally successful are the Gated Recurrent Units (GRUs) \cite{cho2014learning} which, although characterized by a simpler structure with respect to LSTMs, have proven to provide satisfactory performance in a number of applications \cite{bianchi2017recurrent}, see e.g. \cite{bonassi2021nonlinear, lanzetti2019recurrent, mohajerin2019multistep, rehmer2019using}. 
{The architectures listed above are discussed in more detail in Section \ref{sec:RNN}.}

\medskip

In this context, the aim of this paper is to review recent results and to highlight some of the main future challenges related to the applicability and appropriateness of RNN in control-related applications. 
More specifically, focusing on the previously introduced RNN architectures, we discuss how to enforce stability properties during the training procedure, which turn out to be crucial properties for the design of theoretically-sound control strategies, and highlight a number of future research directions aimed at providing RNN with additional properties which could foster their adoption by control engineers.
 Although the analysis is focused on RNN and indirect approaches, we believe that many considerations and results here reported could be also relevant for different structures and control designed methods.
 
It is worth noticing that this paper is not intended to be a complete walkthrough of the use of RNN in systems control domain.
Indeed, the selection of one architecture over another, its design and training, and many tasks known to be fundamental to the success of the identification procedure, such as designing the experiments for the collection of the training data, are well beyond the scope of this paper.
For these aspects, the interested reader is addressed to \cite{schoukens2019nonlinear, forgione2020model, bonassi2021stability}.
\smallskip

%
% Among the many RNN described in the literature, probably the most popular ones are the Neural Nonlinear AutoRegressive eXogenous (NNARX) networks which, in light of their simple structure and training, have been extensively applied for system identification and control, both in academia \cite{levin1993control,levin1995identification, sastry1994memory} and in industry \cite{ali2015artificial,himmelblau2008accounts}.
% The joint use of NNARX models and MPC has been considered in \cite{piche2000nonlinear, hosen2011control,nagy2007model}, with the aim of operating the system in a broader region compared to traditional linear models.

%The third class consists of the Long Short Term Memory (LSTM) networks \cite{hochreiter1997long}, nowadays the most popular RNN in view of their excellent performance in a wide class of problems such as mobile phones, GPS navigators for speech recognition, image analysis, just to name a few.
%Their application to control problems has been recently studied in \cite{wu2019machine, terzi2021learning}.
%The last family we consider here is the one of Gated Recurrent Units (GRU) \cite{mohajerin2019multistep,rehmer2019using} which, although characterized by a simpler structure with respect to LSTM, have proven to provide satisfactory performance in a number of applications \cite{bianchi2017recurrent}.\smallskip

The paper is organized as follows. 
In Section \ref{sec:preliminaries}, the adopted notation and the definition of Input to State (ISS) and Incremental Input to State Stability ($\delta$ISS) are introduced. 
%Such stability notions are used in the following sections to analyze the properties of the considered NN.
 Section \ref{sec:RNN} is devoted to briefly describe the NNARX, ESN, LSTM, and GRU networks, as well as {their training procedure and stability properties}.
 Section \ref{sec:mpc_rNN} deals with the design of closed-loop stable MPC algorithms for the considered RNN architectures. 
 In Section \ref{sec:towards}, open issues related to the use of RNN for control design namely the robustness, \emph{safety verification},  \emph{interpretability} issues, {as well as some of the research trends concerning these topics, are presented.
 Some of the discussed topics are then illustrated on a simulated chemical system in Section \ref{sec:example_physics}.}
 Lastly, open problems and future research directions are shortly discussed in the concluding Section \ref{sec:conclusions}.

\section{Notation and preliminaries} \label{sec:preliminaries}
\noindent \textbf{Notation}
In the paper we adopt the following notation. Given a vector $v$, we denote by $v^{\prime}$ its transpose and by $\| v \|_p$ its $p$-norm.
For conciseness, the value of a time-varying vector $x$ at time $k$ is indicated as $x_k$.
Boldface symbols are used to indicate sequences of vectors, i.e. $\bm{v}_k = \{ v_0, v_1, ..., v_k \}$, whose $\ell_{p, q}$ norm is defined as $\| \bm{v}_k \|_{p, q} = \big\| \,[ \, \| v_0 \|_p, ..., \| v_k \|_p \, ] \, \big\|_q$.
In particular, note that $\| \bm{v}_k \|_{p, \infty} = \max_{t \in \{0, ..., k\}} \| v_t \|_p$.
The Hadamard (i.e. element-wise) product between $u$ and $v$ is indicated by $u \circ v$.
The sigmoidal and hyperbolic tangent activation functions are denoted by $\sigma(x) = \frac{1}{1 + e^{-x}}$ and $\phi(x) = \tanh(x)$, respectively. \rev{Note that when these functions are applied to a vector, they are intended to be applied element-wise.}
\rev{Lastly, given a generic discrete-time system described $x_{k+1} = f(x_k, u_k)$ and $y_k = g(x_k, u_k$), we denote by $x_{k}(\bar{x}, \bm{u}_k)$ the state trajectory at time $k$, obtained initializing the system in $\bar{x}$ and feeding it with the input sequence $\bm{u}_k$, and by $y_k(\bar{x}, \bm{u}_k)$ the corresponding output trajectory.}

\begin{definition} \label{def:kinf_function}
A continuous function $\gamma: {R}_{\geq0}\to{R}_{\geq0}$ is a class $\mathcal{K}$ function if $\gamma(s)>0$ for all $s>0$, it is strictly increasing, and $\gamma(0)=0$. The function $\gamma$ is a class $\mathcal{K}_{\infty}$ function if it is a class $\mathcal{K}$ function and $\gamma(s)\to\infty$ for $s\to\infty$.
\end{definition}

\begin{definition}\label{def:kl_function}
A continuous function $\beta: {R}_{\geq0}\times{Z}_{\geq0}\to{R}_{\geq0}$ is a class $\mathcal{KL}$ function if $\beta(s,k)$ is a class $\mathcal{K}$ function with respect to $s$ for all $k$, it is strictly decreasing in $k$ for all $s>0$, and $\beta(s,k)\to0$ as $k\to\infty$ for all $s>0$.
\end{definition}

\begin{definition}[\cite{jiang2001input}] \label{def:iss}
The dynamical system $x_{k+1}=f(x_k, u_k)$
is Input-to-State Stable (ISS) if there exist functions $\beta( \| \bar{x} \|_2, k) \in \mathcal{KL}$ and $\gamma_u(\| \bm{u}_k \|_{2, \infty} ) \in \mathcal{K}_\infty$ such that for any $k \geq 0$, any initial condition $x_0$,  and any input sequence $\textbf{u}_k$, it holds that
\begin{equation}
\| x_k(\bar{x}, \textbf{u}_k) \|_2 \leq \beta(\| \bar{x} \|_2, k) +\gamma_u(\| \textbf{u}_k \|_{2, \infty} ).
\end{equation}
\end{definition}
It is worth noticing that the ISS of a system implies, for bounded input sequences, the boundedness of the system's state.
\begin{definition}[\cite{bayer2013discrete}]\label{def:deltaiss}
The dynamical system $x_{k+1}=f(x_k,u_k)$
is Incrementally Input-to-State Stable ($\delta$ISS) if there exist functions $\beta( \| \bar{x}_a - \bar{x}_b \|_2, k) \in \mathcal{KL}$ and $\gamma_u(\| \bm{u}_{a,k} - \bm{u}_{b,k} \|_{2, \infty} ) \in \mathcal{K}_\infty$ such that for any $k {\geq 0}$, any pair of initial conditions $\bar{x}_{a}$ and $\bar{x}_b$, and any pair of input sequences $\bm{u}_{a, k}$ and $\bm{u}_{b, k}$, it holds that
\begin{equation} \label{eq:deltaiss:def}
\begin{aligned}
	&\| x_{k}(\bar{x}_a, \bm{u}_{a,k}) -   x_{k}(\bar{x}_b, \bm{u}_{b,k}) \|_2 \\
	& \qquad\qquad \leq \beta(\| \bar{x}_{a} - \bar{x}_{b} \|_2, k) + \gamma(\| \bm{u}_{a, k} - \bm{u}_{b, k} \|_{2, \infty} ).
\end{aligned}
\end{equation}
\end{definition}
The $\delta$ISS property implies that the smaller the distance between two input sequences, the smaller is, asymptotically, the maximum distance between the resulting state trajectories, regardless of the system's initial states.
{This property is stronger than ISS. Indeed, if a system is $\delta$ISS then it is also ISS.}

Among the many direct implications, this property entails that the effects of the initial conditions asymptotically vanish, i.e. the state trajectories are solely determined by the applied input.
{This not only implies that for constant inputs the states converge to unique equilibria, but -- when this property is enjoyed by some model of a dynamical system -- it especially entails that the modeling performances are independent of the model's initialization.}
This latter is a major concern especially when a RNN  is used to identify the plant using input-output data, since the network's states are not only unobserved quantities, but they generally have no physical meaning at all.
In light of the relevance of the $\delta$ISS property, the next section is devoted to discuss how provenly $\delta$ISS RNN models can be trained, while its consequences in control design are discussed in Section \ref{sec:mpc_rNN} and Section \ref{sec:towards}.

\section{Families of Recurrent Neural Networks} \label{sec:RNN}
Four classes of RNN are considered in this paper, namely NNARX, ESN, LSTM, and GRU.
Below we briefly present these architectures assuming, for the sake of simplicity, single-layer structures. 
For LSTM and GRU, multi-layer structures can however be adopted to improve the modeling performances, at the price of a more complex notation and training.

Recurrent networks are stateful NN that can be generally described as a dynamical \rev{MIMO} state-space model,
\begin{equation} \label{eq:rnn:general}
\begin{dcases}
	x_{k+1} = f(x_k,u_k;\Phi)  \\
	y_{k} = g(x_k,u_k; \Phi) 
\end{dcases}	
\end{equation}
where $x \in \mathbb{R}^{n_x}$ is the state vector, $u \in \mathbb{R}^{n_u}$ is the input, $y \in \mathbb{R}^{n_y}$ is the output, and $\Phi$ is the set of parameters, named weights, that are computed during the so-called training procedure, described later in this section.
The structure of functions $f$ and $g$ and the meaning of the state $x$ depend on the selected architecture, as discussed below.

\subsection{NNARX networks}
In NNARX models, it is assumed that the future output $y_{k+1}$ only depends on the past $N$ input and output data.
More specifically, $y_{k+1}$ is computed as a nonlinear regression over the past data,
\begin{equation} \label{eq:rnn:nnarx_model}
	y\sss{k+1} = \eta(y\sss{k}, y\sss{k-1}, ..., y\sss{k-N+1}, u\sss{k}, u\sss{k-1}, ..., u\sss{k-N}; \Phi),
\end{equation}
where $\eta$ is a FFNN parametrized by the weights collected in the set $\Phi$.
Model \eqref{eq:rnn:nnarx_model} can easily be recast in the general form \eqref{eq:rnn:general}.
Specifically, \eqref{eq:rnn:nnarx_model} corresponds to a discrete-time normal canonical form \cite{bonassi2021nnarx}.
Indeed, letting $i \in \{1, ..., N \}$
\begin{equation} \label{eq:rnn:nnarx_states}
	z\sss{i, k} = \left[\begin{array}{c}
		y\sss{k-N+i} \\
		u\sss{k-N-1+i}
	\end{array}\right],
\end{equation}
it can be shown that \eqref{eq:rnn:nnarx_model} is equivalent to
\begin{equation} \label{eq:rnn:nnarx_normal_form}
	\left\{ \begin{array}{l}
		z\sss{1, k+1} = z\sss{2, k} \\
		\quad \vdots \\
		z\sss{N-1,k+1} = z\sss{N, k} \\
		z\sss{N, k+1} = \begin{bmatrix}
			\eta(z\sss{1, k}, z\sss{2, k}, ..., z\sss{N, k}, u\sss{k}; \Phi) \\
			u\sss{k}
		\end{bmatrix} \\
	y\sss{k} = [I\quad 0] \, z\sss{N, k}
	\end{array} \right..
\end{equation}
By defining the state vector as $x\sss{k} = [ z\sss{1, k}^\prime, ..., z\sss{N, k}^\prime]^\prime$, \eqref{eq:rnn:nnarx_normal_form} can be compactly written as
%\begin{subequations} \label{eq:model:nnarx}
\begin{equation}\label{eq:model:statespace}
\begin{dcases} 
	x\sss{k+1} = {A} x\sss{k} + B_{u} u\sss{k} + B_{x} \eta(x\sss{k}, u\sss{k}; \Phi) \\
	y\sss{k} = C x\sss{k}
\end{dcases},
\end{equation}
where $A$, $B_u$, $B_x$, and $C$ are fixed suitable matrices, see \cite{bonassi2021nnarx}.
Concerning the regression function, under mild assumptions any FFNN architecture can be adopted. 
For illustration purposes, the following structure may be assumed, 
\begin{equation}  \label{eq:rnn:nnarx_ffnn}
		\eta(x\sss{k}, u\sss{k}; 	\Phi) = U_0  \psi \big( W_1 u\sss{k} + U_1 x\sss{k} + b_1 \big)  + b_0,
	\end{equation}
where $\psi$ is a Lipschitz continuous activation function, with Lipschitz constant $L_{\psi}$ and satisfying $\psi(0) = 0$.
Hence, the weights of this model are
\begin{equation*}
	\Phi = \{ U_0, b_0, W_1, U_1, b_1 \}. 
\end{equation*}
Lastly, it is worth noticing that NNARX state vector $x_k$ is a collection of past input and output data.
In addition to ensuring the interpretability of network states, this peculiarity allows one to easily employ the NNARX network as prediction model for MPC design.
Indeed, a state observer is not required to operate such model in closed-loop, since, being it a collection of past known data, the actual state is known at any time instant.

\subsection{ESN networks}
Introduced by \cite{jaeger2001echo}, Echo State Networks are dynamic networks characterized by the form
\begin{equation} \label{eq:rnn:esn_statespace}
\begin{dcases}
	 x_{k+1} =  \sigma(W_u u_k+ U x_k + W_y y_k) \\
	y_k =  U_o \, x_k + W_o u_{k-1}
\end{dcases}.
\end{equation}
The peculiarity of this architecture is that the weights $\tilde{\Phi} = \{ U, W_u, W_y \}$ are randomly generated before training, with the only condition that $W_x$ is a sparse and Schur stable matrix.
Then, unlike other NN architectures, these weights are not tuned during the training procedure, so that the state dynamics are fixed once such weights are generated.
The rationale behind this approach is that the state should be a dynamic reservoir, i.e. it should be able to represent any possible stable dynamics by suitably combining these fixed state trajectories.
During the training procedure, the weights $\Phi = \{ U_o, W_o \}$ are tuned so that the ESN represent the plant's dynamics.

What makes ESN interesting is that, under a proper selection of the loss function, the training procedure boils down to a Least Square problem, which greatly simplifies the identification problem, owing to the reliability and efficiency of Least Square algorithms \cite{schoukens2019nonlinear} and to the avoidance of the so-called vanishing gradient problem, which instead plagues other RNN architectures \cite{pascanu2013difficulty}.

\subsection{LSTM networks} \label{sec:RNN:LSTM}
\begin{subequations} \label{eq:rnn:lstm_statespace}
The one of Long Short-Term Memory networks is a widely popular RNN architecture, owing to its performances in learning dynamical systems.
LSTM can be recast in the following state-space form \cite{bonassi2020lstm}:
\begin{equation}
\begin{dcases}
	c_{k+1} = f_k \circ c_k + i_k \circ \phi(W_r u_k + U_r \, h_k + b_r) \\
	h_{k+1} = z_k \circ \phi(c_{k+1}) \\
	y_k = U_o h_k + b_o
\end{dcases},
\end{equation}
where $f_k$, $i_k$, and $z_k$ are the so-called \emph{gates}, that rule the flow of information throughout the network, thus allowing to tackle the vanishing and exploding gradient problem and to retain long-term memory \cite{hochreiter1997long, pascanu2013difficulty}.
More specifically, the gates are described by 
\begin{equation}
\begin{aligned}
	 f_{k}  &= \sigma(W_f u_k + U_f h_k + b_f),\\
	i_{k}  &= \sigma(W_i u_k + U_i  h_k + b_i ),\\
	z_{k} &= \sigma(W_z u_k + U_z h_k + b_z).
\end{aligned}
\end{equation}
\end{subequations}
The LSTM network \eqref{eq:rnn:lstm_statespace} is characterized by the state vector $x_k = [ c_k^{\prime}, h_k^{\prime} ]^{\prime}$, while the set of weights that need to be tuned during the training procedure is
\begin{equation*}
	\Phi=\{ W_f,U_f,b_f,W_i,U_i,b_i,W_r,U_r,b_r,W_z,U_z,b_z, U_o, b_o\}.
\end{equation*} 
Under this notation, the LSTM model \eqref{eq:rnn:lstm_statespace} falls into the generic form \eqref{eq:rnn:general}.
Unlike ESNs, the training of LSTMs is known to be non trivial, due to the large number of weights.

\subsection{GRU networks}
\begin{subequations} \label{eq:rnn:gru_statespace}
A fair tradeoff between architecture's complexity and modeling performances is what characterizes Gated Recurrent Units networks \cite{bianchi2017recurrent}.
Like LSTM, GRU exploit gates to tackle the vanishing and exploding gradient problems, but with a simpler structure to reduce the number of weights.
In particular, the state-space model of GRU reads as follows \cite{bonassi2021stability}
\begin{equation}
\begin{dcases}
	x_{k+1} = z_k\circ x_k + (1 - z_k) \circ \phi ( W_r  u_k + U_r f_k \circ x_k + b_r ) \\
	y_k = U_o x_k + b_o
\end{dcases},
\end{equation}
where $z_k$ and $f_k$ are the gates, described by
\begin{equation}
\begin{aligned}
	z_k &= \sigma \left( W_z  u_k + U_z x_k + b_z \right), \\
    f_k &= \sigma \left( W_f u_k + U_f x_k + b_f \right). 
\end{aligned}	
\end{equation}
\end{subequations}

Denoting by
\begin{equation}
	\Phi=\{W_r,U_r,b_r,W_z,U_z,b_z,W_f,U_f,b_f,U_o,b_o\}
\end{equation}
the weights of the network, \eqref{eq:rnn:gru_statespace} takes the form of \eqref{eq:rnn:general}.

%\begin{remark}
%	Models \eqref{eq:model:statespace} and \eqref{ESNeq} essentially belong to the class of networks denoted as RNN or \emph{Feedback Networks} and studied in the early days of the research on NN in control applications, e.g., in \cite{sontag1992neural}. What makes NNARX of particular interest is the fact that the state actually includes just past inputs and/or outputs, and therefore the use of a state estimator (and its design) is not required. On the other hand, the difference between ENS and ''classic'' Feedback Networks is the set of free parameters. Indeed, as introduced, the peculiarity of ENSs as they are proposed, e.g., in \cite{jaeger2001echo}, is the fact that matrices $W_x$, $W_{u}$, and $W_y$ are essentially given or randomly generated - they are hyperparameters - while, during the training phase, just $W_{out_{1}}$ and $W_{out_{2}}$ are tuned. This greatly simplifyies the identification problems, requiring just a least-square algorithm and allows to overcome the vanishing gradient issue [UN RIFERIMENTO, FABIO? QUESTO REMARKI MI CONVINCE POCO].
%\end{remark}

\subsection{Network training and stability} \label{sec:RNN:training}
Once a RNN architecture is selected, and its hyperparameters -- such as the number of neurons, which is generally related to the state dimensionality -- are chosen, a training procedure must be carried out to tune the weights $\Phi$ so that the network represents an accurate model of the plant.
To this end, input-output sequences collected from the plant are required. 
These data must be ``informative'' enough, meaning that they should be collected through suitably crafted experiments, see \cite{schoukens2019nonlinear}.

A popular approach to RNN training is the so-called Truncated Back-Propagation Through Time (TBPTT) method \cite{bianchi2017recurrent}, which consists in extracting shorter and partially-overlapping input-output subsequences from the input-output data of the experiment.
These subsequences are denoted by $(\bm{u}^{\{i\}}, \bm{y}^{\{i\}}_m)$, where $\bm{u}^{\{ i \}}$ is the input applied to the plant, and $\bm{y}_m^{\{i\}}$ the corresponding measured output, and their length is denoted by $T_s$.
The index $i \in \mathcal{I} = \{ 1, ..., N_s \}$ is used to indicate the subsequences, which are randomly split in a training set $\mathcal{I}_t$, a validation set $\mathcal{I}_v$, and a test set $\mathcal{I}_f$, with $\mathcal{I}_t \cup \mathcal{I}_v \cup \mathcal{I}_f = \mathcal{I}$ and $\mathcal{I}_t \cap \mathcal{I}_v \cap \mathcal{I}_f = \emptyset$.
The network is then trained by minimizing the loss function $L$, defined as the Mean Square Error (MSE) between the RNN prediction and the measured output, i.e.
\begin{equation} \label{eq:rnn:training_problem}
	\min_{\Phi} \big\{ L(\Phi) = \text{MSE}(\mathcal{I}_t; \Phi) \big\}.
\end{equation}
The MSE can be formalized as
\begin{equation}\label{eq:rnn:mse_def}
\scalemath{0.85}{
	\text{MSE}(\mathcal{I}_\alpha; \Phi) = \frac{1}{\lvert \mathcal{I}_\alpha \lvert (T_s - T_w)} \sum_{i\in \mathcal{I}_\alpha} \sum_{k=Tw}^{T_s} \big\| y_k(x_0, \bm{u}^{\{i\}}; \Phi) - y_{m,k}^{\{i\}} \big\|_2^2},   
\end{equation}
 where $\lvert \mathcal{I}_\alpha \lvert$ denotes the number of subsequences in the set $\mathcal{I}_\alpha$, with $\alpha = t, v, f$, and $y_k(x_0, \bm{u}^{\{i\}}; \Phi)$ indicates the output of the RNN \eqref{eq:rnn:general} initialized in the random state $x_0$ and fed by the input sequence $\bm{u}^{\{i\}}$. 
Because of the random initialization, the first $T_w$ steps, known as washout period, are generally discarded \cite{bianchi2017recurrent, bonassi2021stability}.
The training problem \eqref{eq:rnn:training_problem} can be solved by many gradient-based algorithms, such as Adam and RMSProp \cite{bianchi2017recurrent}, halting the procedure e.g. when the MSE on the validation set, $\text{MSE}(\mathcal{I}_v; \Phi)$, stops improving.
Lastly, the modeling performances are assessed on the independent test set $\mathcal{I}_f$. \medskip

At this stage, except for the input-output data, no knowledge of the plant has been exploited during the training procedure.
However, it may be the case that the plant is known to enjoy ISS- or $\delta$ISS-like stability properties, that can be e.g. numerically inferred from the data, or qualitatively deduced from the plant behavior.
One may wonder if it is somehow possible to train a RNN model provenly enjoying such properties, which would allow not only a ``consistent'' model, but also a relevant theoretical tool that can be spent for control design purposes.

\begin{proposition}
	\label{prop:stability}
Under suitable conditions on their weights $\Phi$, NNARXs \cite{bonassi2021nnarx}, ESNs \cite{armenio2019model}, LSTMs \cite{terzi2021learning}, and GRUs \cite{bonassi2021stability} are guaranteed to be ISS and $\delta$ISS.
These conditions, synthetically reported in the Appendix, can be generally regarded as  nonlinear inequalities on the weights of the network, denoted by
\begin{equation} \label{eq:rnn:stability_condition}
	\nu(\Phi) < 0.
\end{equation}
\end{proposition}

In other words, if the inequality \eqref{eq:rnn:stability_condition} holds, the RNN model is guaranteed to be ISS and/or $\delta$ISS.
Hence, if a stable RNN is sought, condition \eqref{eq:rnn:stability_condition} can be easily enforced during the training procedure by penalizing its violation in the loss function, i.e. by taking
\begin{equation} \label{eq:rnn:stable_loss}
	L(\Phi) = \text{MSE}(\mathcal{I}_t; \Phi) + \rho \big( \nu(\Phi) \big),
\end{equation}
where $\rho(\nu)$ is a \rev{regularization term that increases with $\nu(\Phi)$. }
\rev{For example, if $\nu(\Phi)$ is a scalar, piece\-wise-linear functions can be adopted \cite{bonassi2021stability}, i.e.
\begin{equation}
	\rho(\nu(\Phi)) = \omega_{\scriptscriptstyle -} \min(0, \nu(\Phi)) + \omega_{\scriptscriptstyle +} \max(0, \nu(\Phi)),
\end{equation}
with $0 < \omega_{\scriptscriptstyle -} \ll \omega_{\scriptscriptstyle +}$. This term allows to steer $\nu(\Phi)$ to negative values, enforcing the stability of the network (see Proposition \ref{prop:stability}).
In general $\nu(\Phi)$ is a vector, in which case $\rho(\nu(\Phi))$ can be taken as the component-wise sum of a strictly increasing function applied element-wise on $\nu(\Phi)$. 
}
Note that even in presence of this stability enforcing term, the overall training procedure is the same, except for the fact that the stopping rule should also account whether or not the inequality \eqref{eq:rnn:stability_condition} is fulfilled.

\rev{
\begin{remark}
	Delving into the details about the impact of this stability-enforcing term onto the training procedure is out of the scope of this paper, for which the interested reader is addressed to \cite{bonassi2021stability}.
	However, let us point out that enforcing the model's ISS or $\delta$ISS is reasonable only if the system to be learned displays an analogous property.
	This rules out, e.g., systems with locally unstable equilibria.
	Enforcing the stability condition when learning such unsuitable systems may lead to poor models. 
	It is hence  advisable to assess that the system's trajectories display stability-like properties before setting up the training procedure. 
\end{remark}
}

\section{Model Predictive Control design for models learned by RNN}\label{sec:mpc_rNN}
According to the indirect data-driven control design para\-digm, once a model of the system is identified, a controller is synthesized using any model-based approach.
Let us consider the case in which such model corresponds to a RNN, trained according to the guidelines discussed above, and that the modeling performances on the independent test set turned out to be satisfactory.
This RNN model is thus a good candidate for controller synthesis.

A popular control strategy for RNN models is nonlinear MPC, see e.g. \cite{lanzetti2019recurrent}, which allows to exploit the long-term prediction capabilities of these models while fulfilling input constraints.
In this framework, the main challenges are (\emph{i}) designing an MPC control law which guarantees nominal closed-loop stability and (\emph{ii}) managing the uncertainty of the model.
These two issues are discussed in the following.

\subsection{MPC design in the nominal case} \label{sec:mpc_nominal}
\begin{figure}
	\centering
	\includegraphics[width=0.8\linewidth]{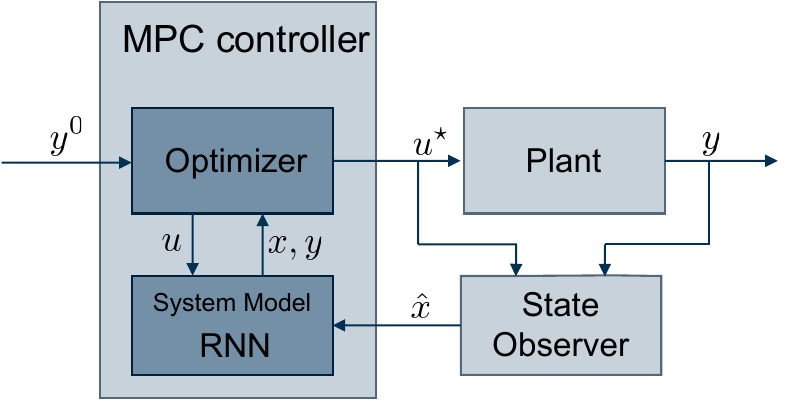}
	\caption{Schematic of the output-feedback controller based on MPC.}
	\label{fig:mpc_architecture}
\end{figure}

The design of an MPC law guaranteeing nominal stability relies on the so-called Certainty Equivalence Principle (CEP).
According to this principle, which is standardly evoked by indirect design approaches, the model and the plant are assumed to match, and any plant-model mismatch is assumed to be associated to a mismatch between the plant's and model's initial states.
Under this principle, an MPC law is designed for the available model, for which well established algorithms guaranteeing recursive feasibility and closed-loop stability exist, see e.g. \cite{rawlings2017model}.
Typically, the resulting output feedback controller, depicted in Figure \ref{fig:mpc_architecture}, is made by an observer, which estimates the state $x_k$ of model \eqref{eq:rnn:general} from output measurements, and a {finite-horizon control} optimization problem which relies on such state estimation.
An exception to this is represented by NNARX models, which do not require state observers, as their state vector solely consists of known past input and output data.

Let us assume, for the sake of generality, that a state observer is required by the controller. 
In this context, the stability properties of the RNN model are beneficial.
In particular, the $\delta$ISS has been shown to entail the possibility to easily design state observers with estimation convergence guarantees, see e.g. \cite{terzi2021learning, bonassi2021nonlinear, armenio2019model, alhajeri2021machine}, where Luenberger-type observers have been proposed for the main RNN architectures.
These observers generally take the following form
\begin{equation} \label{eq:mpc:obsv_general}
	\hat{x}_{k+1} = f_o(\hat{x}_k, u_k, y_k; \tilde{\Phi}),
\end{equation}
where $\tilde{\Phi} = \{ \Phi^\star, \Phi_o \}$ is the set of parameters, including the weights of the trained RNN model, $\Phi^\star$, and the gains of the observer, $\Phi_o$.
For suitable observer structures $f_o$, conditions on the observer gains $\Phi_o$ have been provided to guarantee that $\hat{x}_k$ converges to $x_k$.
It is nonetheless worth noticing that, in view of the model's $\delta$ISS, an open-loop observer $\hat{x}_{k+1} = f(\hat{x}_k, u_k;\Phi)$, replicating the model dynamics \eqref{eq:rnn:general}, would enjoy nominal convergence.

Lastly, state-feedback nonlinear MPC laws with recursive feasibility and closed-loop stability guarantees can be designed exploiting the model's $\delta$ISS, see \cite{terzi2021learning, armenio2019model}.

%More in general, concerning the design of observers with asymptotic convergence properties, note that for system \eqref{eq:rnn:general}, letting $\hat{x}$ be the state estimate, the trivial open-loop observer
%\begin{eqnarray}\label{obs}
%  \hat{x}_{k+1} &=& f(\hat{x}_k,u_k,\Phi)
%\end{eqnarray}
%guarantees the asymptotic convergence $ \hat{x} \rightarrow x$ in view of the $\delta$ISS property.
%More sophisticated observers, together with stabilizing control laws, have been developed in \cite{armenio2019model} for ESN  and in
%\cite{terzi2021learning} for LSTM, where the tracking problem has been considered. The use of GRU in the design of MPC schemes with integrators has been studied in \cite{bonassi2021nonlinear}.

\subsection{MPC design in the case of modeling errors}
Unfortunately, in most cases the assumption of no plant-model mismatch is very strong and not acceptable.
A straightforward, yet ineradicable, reason for such mismatch is that, as for any black-box identification technique, there is no direct correspondence between the RNN's and the plant's states, with the former typically being higher-dimensional than the latter. 

In light of the plant-model mismatch, one can rely on a robust control law.
To this regard, if both the plant and the model are input-output stable (which property is, for example, implied by ISS and $\delta$ISS), one can assume that the plant is described by the RNN state equations \eqref{eq:rnn:general}, with an additive output disturbance, i.e. $y_{m,k} = y_k + d_k$.
The term $d_k \in \mathcal{D}$ represents the plant-model mismatch, whose boundedness is guaranteed by the plant's and model's input-output stability.
The set $\mathcal{D}$ can be estimated from the data, e.g. using the Scenario Approach, as discussed in \cite{terzi2021learning}.
Having a non-conservative bound of $\mathcal{D}$ and a set of admissible plant's outputs $\mathcal{Y}_m$, one can design a nominal MPC law for the model, with an additional constraint on the model's output, i.e. $y_k \in \mathcal{Y}_m \ominus \mathcal{D}$.

While this guarantees that the plant's output is admissible for any realization of the mismatch $d_k$, the feasibility of this approach mainly depends on the conservativeness of the bound on $\mathcal{D}$.
Besides, a more subtle problem is that {this output disturbance must be accounted when proving the convergence of the state observer leading, potentially, to the impossibility to attain any guarantee.
Proving the nominal closed-loop stability entailed by the state-feedback MPC law may thus be challenging.}

In order to foster a wide and robust use of RNN for MPC design, research activities should thus focus not only on how to profitably bound $\mathcal{D}$, but also on how to ensure the observer's convergence and the closed-loop stability in the presence of plant-model mismatch.

% For the above reason, a robust MPC approach must be taken. To this regard, it is reasonable to assume that the real system is described by the I/O model $y_{r_{k}}=S_r(u_k)$, while the NN model is $y_{m_{k}}=S_m(u_k)$ with $y_{m_{k}}=y_{r_{k}}+d_k$, where $d$ is a suitable disturbance. The set of values $D$ achievable by the disturbance can be estimated in different ways starting from the available data, such as the scenario approach, as discussed in \cite{terzi2021learning}. Once the set $D$ has been estimated, and denoting by $Y_r$ the set of admissible outputs for $y_r$, a robust MPC problem can be stated for the nominal model $y_{m_{k}}=S_m(u_k)$ with the additional constraint $y_m\in Y_r \ominus D$. Different possible procedures for the estimation of $D$ can be used and influence the conservativeness of the approach. Deeper insight in this topic is required for a safe and wide use of RNN in control. PROBABILMENTE TROPPO SINTETICO

\section{Towards safe and interpretable RNN for dynamical systems modeling}\label{sec:towards}
Owing to their modeling capabilities, RNN have the potential to play an ever increasing role in the design, management, and control of dynamical systems, even of safety-critical ones, such as in the automotive and aerospace industries \cite{tan2000neural, kurd2005using}.
In order for RNN to be legitimately used in safety critical applications, however, adequate properties have to be certified, in terms of verifiability, robustness, and interpretability, see \cite{ruan2018reachability}. 
In the following, recent results regarding these issues are outlined.

\subsection{Verifiability and robustness of RNN models} \label{sec:towards:verifiability}
When using an identified black-box model for control design, especially if such model is a RNN, a common requirement is that of \emph{generalization}.
A model generalizes well if it produces meaningful and consistent predictions even for data not included in the training set.
While the generalization capabilities of RNN are generally assessed on the so-called independent test set, see Section \ref{sec:RNN:training}, \rev{recent contributions have formulated bounds on the generalization error of some single-layer RNN architectures \cite{chen2020generalization, wu2021statistical}. 
Closely related to the issue of generalization, when a RNN model is used to learn a stable plant,} one may want to certify that (\emph{i}) adversarial perturbations, i.e. small variations of the inputs, do not produce catastrophic changes of the outputs, and that (\emph{ii}) for any possible input sequence the model's outputs always lie in a set of physically-meaningful values.
For example, if a RNN is used to learn the model of a water tank where the input is a controllable inlet flow rate, one may want to ensure that the predicted water level does not undergo abrupt changes for small variations of the inlet flow, and that it always lies in a set of admissible values, e.g. ranging from zero (empty tank) to the maximum level (saturated tank). \medskip

\noindent \emph{i. Robustness} 

{The above-mentioned issue (\emph{i}) corresponds} to the requirement of robustness against input perturbations, also called adversarial attacks or matched disturbances.
Ideally, {to certify this property one should test the RNN model's response} to a sufficiently large amount of perturbed input trajectories \cite{hazan2017adversarial, guo2021rnn}.

On the other hand, being able to show the model's $\delta$ISS allows, by definition, to assess its robustness.
Indeed, recalling Definition \ref{def:deltaiss}, the function $\gamma$ can be used to bound the response of the system to adversarial attacks.
Denoting by $\bm{\delta u}_k$ the input perturbation, from \eqref{eq:deltaiss:def} it follows that the resulting deviation of the state trajectory is bounded as $\| \delta x_k \|_2 \leq \gamma( \| \bm{\delta u}_k \|_{2, \infty})$.
It is worth noticing that function $\gamma$ can be conservatively computed from the weights of the network, see \cite{terzi2021learning, bonassi2021nnarx, armenio2019model, bonassi2021stability}.
However, if the RNN model is $\delta$ISS, less conservative bounds on $\gamma$ can be computed numerically on a sufficiently large set of model trajectories. 
To this end, approaches similar to that presented in \cite{tempo2013randomized} can be adopted.
Lastly, note that alternative strategy to certify the robustness of RNN models is to bound their Lipschitz constant.
The available approaches, however, are currently limited to FFNN, see e.g. \cite{fazlyab2019efficient}. \medskip

\noindent \emph{ii. Safety verification} 

The second issue, known as {safety verification}, consists of certifying that the output reachable set of the RNN model, given bounded inputs and initial conditions, is consistent with the possible values taken by the plant's outputs.
This allows to conclude, for example, that the model's outputs remain meaningful even when evaluated with input sequences not included in the training set.

Unfortunately, the estimation of the output reachable set is difficult to perform for generic nonlinear dynamical systems, and most of the results nowadays available have been developed for FFNN, see \cite{ruan2018reachability, xiang2018output, dvijotham2018dual}.
In principle, these approaches are applied to RNN by ``unrolling'' them, i.e. transforming a  $k$-step RNN simulation in a sequence of $k$ suitable FFNN, see \cite{bianchi2017recurrent} for more details, and then applying the verification algorithms available for FFNN \cite{weng2019popqorn}. 
An interesting method to turn the RNN verification into a non-recurrent problem has been proposed in  \cite{Jacoby}, while a direct analysis of RNN and their generalization properties has been reported in \cite{chen2020generalization, wu2021statistical}, \rev{where bounds on the generalization error have been derived for some single-layer RNN architectures}. \smallskip

An alternative method for the estimation of the output reachable set is based on the so-called Scenario Approach \cite{campi2018introduction}, see the recent contributions \cite{terzi2021learning, bonassi2020lstm, wang2020scenario}, which allows to bound, {with some prescribed confidence}, the output reachable set.
To apply this approach, the boundedness of the output reachable set must be assessed, e.g. by proving the ISS or $\delta$ISS of the RNN model.
In the following, an outline of the probabilistic safety verification procedure based on the Scenario Approach is reported. The interested reader is addressed to \cite{bonassi2020lstm} and \cite{hewing2019scenario} for the details.

Assume that the initial state $\bar{x}$ of the RNN is a random variable extracted from a set $\mathcal{X}_0$ and characterized by some probability measure $\mathbb{P}_{x}$.
Consider a time horizon $K$, and a class $\bm{\mathcal{U}}_K$ of bounded input sequences $ \bm{u}_K = \{ u_0, ..., u_K \}$. %, such that $u_t \in \bm{\mathcal{U}}_k$ for all $t \in  \{ 0, ..., k \}$. 
The class of inputs $\bm{\mathcal{U}}_K$ is assumed to be characterized by some probability measure $\mathbb{P}_u$.
Let $\bm{y}_K(\bar{x}, \bm{u}_K)$ be output of the RNN model \eqref{eq:rnn:general}, initialized in the random initial state $\bar{x} \in \mathcal{X}_0$ and fed by the random input sequence $\bm{u}_K$ drawn from $\bm{\mathcal{U}}_K$ according to $\mathbb{P}_u$.
Then, a bound of the output reachable set may be defined as the smallest set $\mathcal{Y}$, containing, for any possible $\bar{x} \in \mathcal{X}_0$ and $\bm{u}_K \in \bm{\mathcal{U}}_K$, the output $y_k(\bar{x}, \bm{u}_K)$ at any time instant $k \in \{ 0, ..., K \}$.
More specifically, the set $\mathcal{Y}$ may be defined as a suitable convex set $\tilde{\mathcal{Y}}$ scaled by a coefficient $\rho_y$ \cite{hewing2019scenario}, i.e. $\mathcal{Y} = \rho_y \tilde{\mathcal{Y}}$.
Under this definition, bounding the output reachable set boils down to finding the smallest possible $\rho_y$ satisfying this condition.
Note that, at this stage, the formulated problem is infinite-dimensional, and hence it can not be solved. 

This problem can be tackled by applying the Scenario Approach, which allows to relax the infinite-dimensional problem into a deterministic one. 
To this end, one needs to generate a number $S$ of scenarios, each corresponding to an independent sample of the uncertain variables $\bar{\mathrm{x}}^{(s)}$ and $\mathbf{u}_K^{(s)}$, with $s \in \{1, ..., S \}$, drawn from the respective sets according to the associated probability density functions. 
Then, the tightest bound of the output reachable set is determined as
\begin{equation}
\begin{aligned}
	\rho_y^\star = \arg\min_{\rho_y} & \quad \rho_y  &\\
	\text{s.t.} & \quad \rho_y \geq 0 \\
	& \quad y_k(\bar{\mathrm{x}}^{(s)}, \mathbf{u}_k^{(s)}) \in \rho_y \, \tilde{\mathcal{Y}} \quad & \forall k \in \{ 0, ..., K \}, \\
	&&\forall s \in \{ 1,...,S\}.
\end{aligned}
\end{equation}
Defining $\varepsilon_S \in (0, 1)$ as the violation probability, and $1 - \beta_S \in (0, 1)$ as the confidence of such measure, the Scenario Approach guarantees that if
\begin{equation*}
 S \geq \frac{2}{\varepsilon_S} \left( \ln\frac{1}{\beta_S} + 1 \right)
\end{equation*}
then, with confidence $1 - \beta_S$, the probability to draw $\bar{\mathrm{x}} \in \mathcal{X}_0$ and $\mathbf{u}_K \in \mathcal{U}_K$ such that, at some time instant $k \in \{ 0, ..., K \}$, it holds that $ y_k(\bar{\mathrm{x}}, \mathbf{u}_k ) \notin \rho_y^\star \, \tilde{\mathcal{Y}}$ is lower than $\varepsilon_S$.
Thus $\mathcal{Y} = \rho_y^\star \, \tilde{\mathcal{Y}}$ is a probabilistic estimation of the output reachable set, associated to a maximum violation $\varepsilon_S$, with confidence $1 - \beta_S$.
Lastly, this set is compared to the known safe output set, which is e.g. determined by physical constraints of the plant, to certify the safety of the RNN model.

\begin{remark}
This approach for safety verification is theoretically viable only if both the plant under control and its RNN model enjoy the ISS property. 
In fact, as discussed in \cite{reviewISS_Prieur}, ISS is simultaneously equivalent to the Continuity at the Equilibrium Point, to the existence of a Uniform Asymptotic Gain, and to the Boundedness of the Reachability Set. 
Hence, in the context of safety verification, the ISS and $\delta$ISS conditions mentioned in Proposition \ref{prop:stability} provide a solid theoretical foundation for the applicability of the described approach.
\end{remark}

\subsection{Interpretability of RNN models} \label{sec:towards:interpretability}
Another fundamental property to foster the use of RNN as models of dynamical systems is that of \emph{interpretability}.
According to \cite{zhang2021survey}, interpretability is ``the ability to provide explanations in understandable terms to a human''.
Owing to its relevance, interpretability is a currently popular research topic in the Machine Learning community; surveys on recent contributions in the field can be found in \cite{zhang2021survey}, \cite{fan2021interpretability}.

In the context of RNN applied to scientific and engineering domains, such as earth systems, climate science, quantum chemistry, biological sciences, and control, the requirement of interpretability mainly corresponds to the need to guarantee consistency between the RNN model and the known underlying physical laws \cite{willard2020integrating}.
Such physical laws might, for example, imply that some outputs fulfill the mass conservation principle, or that they are positive or enjoy some monotonicity properties.

The new branch of Machine Learning techniques that try to merge physical knowledge into RNN modeling takes various names, such as Theory-Guided Data Science \cite{karpatne2017theory}, eXplainable Artificial Intelligence \cite{adadi2018peeking}, or merely Physics-Based Modeling \cite{willard2020integrating, thuerey2021physics}.
A common denominator of these approaches is the attempt to overcome the limitations of black-box modeling, and to shape the adopted machine learning technique according to a grey-box modeling criterion driven by the available physical knowledge of the system.
Depending on the case at hand, such grey-box approaches consist of shaping the structure of the RNN model in specific ways or using, during the training procedure, a suitable loss function, so as to impose consistency with the physical knowledge of the system.
Contributions in this direction have been proposed by many authors, see e.g. the definition of Semi-Empirical NN \cite{egorchev2018semi}, the use of canonical forms \cite{dreyfus1998canonical}, or the methods described in \cite{oussar2001gray} and \cite{beucler2021enforcing}.
As for control, a notable and almost unique contribution to this emerging field has been presented in \cite{wu2019machine}, where a methodological approach has been devised and applied to a simulated chemical process.
In the following, the main approaches towards physics-based RNN models are briefly outlined.

\subsubsection{Physics-based structure design} \label{sec:towards:model_design}

One of the main ways to move from purely black-box NN models to physics-based ones is to embed the physical knowledge of the system by suitably designing the structure of the NN model, see  \cite{karpatne2017theory, willard2020integrating}.
Some of the proposed structures are listed below. \medskip

\noindent \emph{i. Models with known and measurable states} 

When the states of the system to be identified are known and measurable, a simple -- yet effective -- strategy may rely on FFNN to learn the increments of the discretized state variables. 
More specifically, consider the underlying plant's equations to be described by the unknown continuous time model
\begin{equation*}
  \dot{x}_{c}(t)=\varphi_c(x_c(t),u(t)),
\end{equation*}
where the state variables are fully measured, and $t$ is the continuous time index.
By discretizing this system with sampling time $\tau$ and any explicit method, such as Forward Euler or Explicit Runge-Kutta, the resulting model is
\begin{equation*}
  x_{k+1} - x_k = \tau \varphi_d(x_k,u_k).
\end{equation*}
Based on this model, a FFNN can thus be trained to describe the state increment $\varphi_d(x_k,u_k)$, leading to a strategy closely resembling the popular ResNet \cite{he2016deep} and ODE-NN \cite{chen2018neural} architectures. 
An example of application of this approach is \cite{pozzoli2020tustin}, where it has been applied to model a fourth order mechanical system, with states being the  measured angles and the estimated angular speeds.
\rev{A similar strategy can also be adopted when a first-principle (yet not sufficiently accurate) model of the system, denoted as $\bar{\varphi}_d(x_k, u_k)$,  is available. 
In this case, one can use a NN to learn the model residual \cite{wu2020process}, i.e. the function $\Delta \varphi_d(x_k, u_k)$ such that
\begin{equation*}
	x_{k+1} - x_k = \tau \bar{\varphi}_d(x_k, u_k) + \tau \Delta \varphi_d(x_k, u_k).
\end{equation*}
}
 \medskip

\noindent \rev{\emph{ii. Models with known states and structure parametrized by NN}}

\rev{Another case is that of physical systems that enjoy a model characterized by measurable states and a known structure, which however depends on parameters that are unknown and varying based on the system's operating conditions, i.e.
\begin{equation*}
	x_{k+1} = \varphi(x_k, u_k, \check{\Phi}(x_k, u_k)),
\end{equation*}
where $\check{\Phi}(x_k, u_k)$ denotes the model's parameters, assumed to be a function of the state and input.
A common choice in this case is to learn $\check{\Phi}(x_k, u_k)$ using a NN, generally a FFNN, that is trained by minimizing the model's simulation error \cite{hosen2011control, alhajeri2021machine}.
}
\medskip

\noindent \emph{iii. Models with known relationships among variables} 

Outputs describing certain physical quantities are typically subject to constraints that affect them individually, such as positivity, monotonicity \rev{and range bounds}, or that bind them together, such as a zero-sum constraints coming from mass conservation.
Therefore, one may tailor the structure of the RNN model to ensure that the output variables fulfill the established constraints, thus assigning a physical meaning to the output variables themselves.
\rev{In this regard, a typical approach is to properly design the output transformation associated with such outputs.}
A valuable contribution to this approach is provided by \cite{daw2020physics}, where the authors show how a RNN can be designed to model the water temperature profile at different depths of a lake.
Moving from the ideal physical fact that the water temperature should monotonically decrease with the depth, a monotonicity-preserving structure which makes use of intermediate physical variables is designed based on LSTM.
Other well-documented examples of this approach are reported in \cite{willard2020integrating}.
Notably, similar design arguments can be easily applied to many systems, such as trays of binary distillation columns \cite{mertens2018monotonic}, where concentrations and temperatures profiles follow monotonic trends, and chains of chemical reactors, where masses, flow rates, and concentrations obey to mass conservation principle.
%These ideas are discussed on a numerical example in Section \ref{sec:example_physics}.
\medskip

\noindent \emph{iv. Models reflecting the plant's block structure}

Many complex processes can be decomposed in sparsely-connected subsystems, i.e. subsystems whose dynamics are directly influenced only by the neighboring ones.
When the overall system's structure is known, and those coupling variables that describe the subsystems' mutual influences are measured, a RNN can be designed and trained mimicking the architecture of the plant.
\rev{Thus, non-physical input-output connections are ruled out by adopting sparsely-connected RNN, see e.g. \cite{wu2020process, alhajeri2022process}}.
Leveraging this knowledge of the system structure allows one to obtain not only a model that is more reliable, but also less prone to overfitting.
This is typically achieved with a significantly faster convergence rate of the training procedure \cite{alhajeri2022process}.
A numerical example of application of the above approaches is described in Section \ref{sec:example_physics}.

\subsubsection{Physics-guided cost function} \label{sec:towards:cost_function}
Rather than hard-coding the consistency to physical laws via structure selection, one can also embed physical knowledge by including suitable terms in the training loss function, with the goal of encouraging certain behaviors of the resulting RNN model \cite{willard2020integrating}.
This approach, which is particularly useful when the conditions for physical consistency cannot be easily encoded via structure selection, is, in a sense, reminiscent of constraint relaxation.
Seen through these lenses, the idea of enforcing the RNN model's ISS and $\delta$ISS properties by including their conditions in the training loss function, discussed Section \ref{sec:RNN:training}, can be considered as an application of Physics-guided cost function design, which aims to ensure consistency with the prior knowledge that the plant enjoys the same stability properties \cite{bonassi2021stability}.
On the other hand, as one may expect, the main challenge of this approach boils down to suitably choosing how these additional terms of the loss function are weighted.
Finding a trade-off between the modeling performances and the fulfillment of the physical condition may be non-trivial and usually requires trial-and-error tuning procedures.

\section{Numerical Example: learning a physics-based RNN model of a chemical plant}\label{sec:example_physics}
\begin{figure}[t]
	\centering
	\includegraphics[width=\linewidth, clip, trim=1mm 0mm 2mm 0mm]{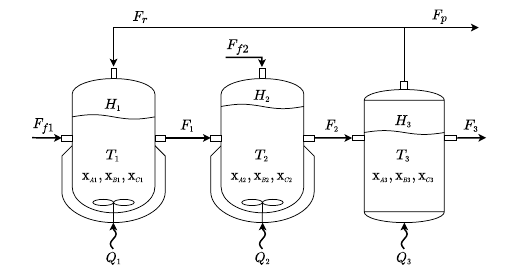}
	\centering
	\caption{Two reactors in series with separator and recycle}
	\label{fig:benchmark_system}
\end{figure}

In this section, we illustrate how a physics-based RNN can be designed for a popular chemical plant benchmark system.
This system, described in \cite{stewart2010} and illustrated in Figure \ref{fig:benchmark_system}, consist of two reactors and one separator.
In the two reactors ($i=1$ and $i=2$), the reactant liquid $A$ is converted into the product $B$ and side-product $C$.
This reaction is controlled by adjusting the flow rate of product $A$ to the two reactors, indicated by $F_{f1}$ and $F_{f2}$, and the external heat $Q_1$ and $Q_2$. 
Then, the mixture enters the separator ($i=3$), where additional heat $Q_3$ is supplied.
The distillate, which is partially fed back to the first reactor via the flow rate $F_r$, is the final product, with constant flow rate $F_p$.
For each subsystem $i \in \{ 1, 2, 3 \}$ the state of the mixture is described by the temperature $T_i$, the level $H_i$, and the concentrations of the three components, denoted by $\mathrm{x}_{Ai}$, $\mathrm{x}_{Bi}$, and $\mathrm{x}_{Ci}$, while the flow rates between the vessels are denoted by $F_i$.

The physical model of the process, derived by mass and energy balance equations, is a nonlinear dynamical system characterized by $m=6$ inputs and $n=12$ states, i.e.\begin{equation}
\scalemath{0.88}{
	\begin{aligned}
u &= [Q_{1},\; Q_{2},\; Q_{3},\; F_{f1},\; F_{f2},  F_{R}, ]^\prime, \\
x &= [H_{1},\; \mathrm{x}_{A1},\; \mathrm{x}_{B1},\; T_{1},\; H_{2},\; \mathrm{x}_{A2},\; \mathrm{x}_{B2},\; T_{2},\; H_{3},\; \mathrm{x}_{A3},\; \mathrm{x}_{B3},\; T_{3}]^\prime. 
\end{aligned}}
\end{equation}
Note that the state vector of the model does include the concentration of the side-product $C$, i.e. $\mathrm{x}_{C i}$, since the latter can be derived by the following relationship among the concentrations: 
\begin{equation} \label{eq:example:concentrations}
	\mathrm{x}_{A i} + \mathrm{x}_{B i} + \mathrm{x}_{C i } = 1.
\end{equation}
The states are assumed to be measurable, so the output coincides with the states, i.e. $y=x$. 
The equations of the model, alongside its parameters, are reported in \cite{stewart2010}, while the input and state variables are summarized in Table \ref{tab:benchmark_symbols}.

\begin{figure}[t] 
\includegraphics[width=0.9\linewidth]{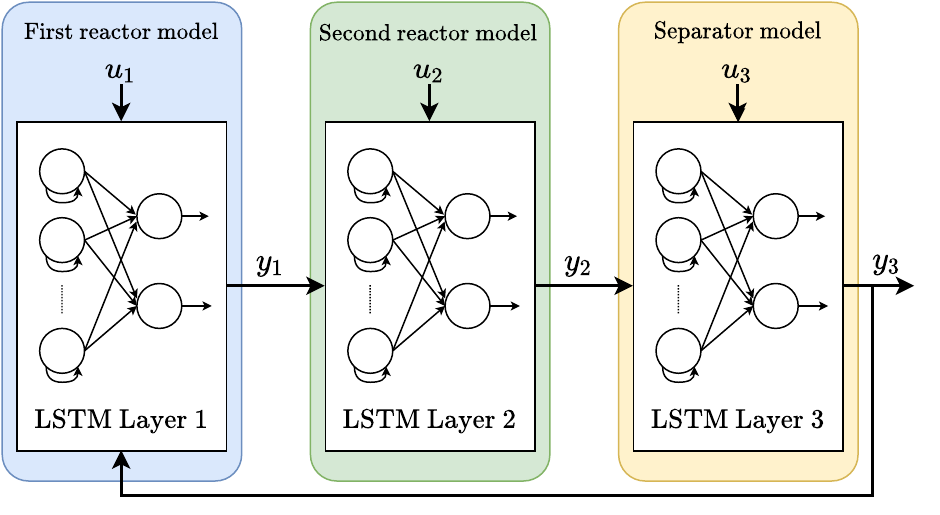}
\centering
\caption{Structure of the physics-based LSTM  used to identify the plant.}
\label{fig:physics}
\end{figure}

\begin{table}
\centering
\resizebox{\columnwidth}{!}{
\begin{tabular}{@{}lllll@{}}
\toprule
Symbol & Unit &  Definition  \\ \midrule
$F_{r}$, $F_{p}$ & $kg/s$  & Recycle and final product flow rates   \\
$F_{f1}$, $F_{f2}$& $kg/s$ & Input flow of reactant $A$ to the reactors    \\
$Q_{i}$& $kj/s$ & External heating supplied to vessel $i$  \\
$H_{i}$& $m$ & Liquid's level in vessel $i$   \\
$T_{i}$& $K$ & Liquid's temperature  in  vessel $i$ \\
$\mathrm{x}_{Ai}$ & $wt\%$  & Concentration of reactant $A$ in vessel $i$   \\
$\mathrm{x}_{Bi}$ & $wt\%$  & Concentration of product $B$ in vessel $i$  \\
$\mathrm{x}_{Ci}$ & $wt\%$  & Concentration of side-product $C$ in vessel $i$  \\
$F_{i}$& $kg/s$  & Outlet flow rate of vessel $i$ \\
\bottomrule
\end{tabular}}
\caption{Summary of plant's input and state variables.}
\label{tab:benchmark_symbols}
\end{table}

% Decentralized structure
\begin{subequations}
As discussed in \cite{stewart2010}, this model notably enjoys a sparse block structure.
Denoting by $x_{i}$ the states associated to vessel $i \in \{ 1, 2, 3 \}$, where
\begin{equation}
	x_{i} = [ H_i, \;\; \mathrm{x}_{A i}, \;\; \mathrm{x}_{B i}, \;\; T_i ]^\prime, 
\end{equation}
and by $y_i = x_i$ its outputs, one can notice that the dynamics of each vessel are only affected by a subset of the input $u_i$, where
\begin{equation}
		u_1 = [ Q_1, \; F_{f_1}]^\prime, \quad 
		u_2 = [ Q_2, \; F_{f_2}]^\prime, \quad
	    u_3 = [ Q_3, \; F_{r}]^\prime, \\
\end{equation}
\end{subequations}
and by the states of incoming flow rates.
As discussed in Section \ref{sec:towards:model_design}, this peculiar structure, characterized by sparsely connected subsystems, can be leveraged to design a physics-based RNN.
Specifically, the structure depicted in Figure \ref{fig:physics} has been adopted, where three distinct LSTM networks have been used to identify the dynamics of the three subsystems.
Hence, for example, LSTM $1$ is intended to learn the dynamics of vessel $1$, given the inputs $\tilde{u}_1 = [ u_1, \; y_3 ]^\prime$, containing the control action $u_1$ and $y_3$, i.e. the output of the LSTM modeling vessel $3$.

Another physical argument that has been leveraged to ensure the consistency of the physics-based network is that, for each subsystem, the output components  $\mathrm{x}_{A i}$ and $\mathrm{x}_{B i}$ must lie, by definition, in the interval $(0, 1)$.
This can be attained by adopting a sigmoid activation function for those output components.
Thus, each subsystem $i \in \{ 1, 2, 3 \}$ is learned by an LSTM network described by the following equations
\begin{equation}
\begin{dcases}
	x_{i, k+1} = f(x_{i,k}, \tilde{u}_{i,k}; \Phi_i) \\
	y_{i, k} =  \Sigma \odot g(x_{i,k}, \tilde{u}_{i,k}; \Phi_i)
\end{dcases},
\end{equation}
where the function $f$ and $g$ are those described in Section \ref{sec:RNN:LSTM}, and $\Sigma \, \odot \, g$ indicates the function composition between the output transformation $g$ and the vector of activation functions
 $\Sigma = [ \imath, \; \sigma, \; \sigma, \; \imath ]^\prime$, $\imath$ being the identity function.

Moreover, a further physical argument is used to ensure the physical consistency of the model.
Specifically, for each vessel the concentrations should satisfy
\begin{equation} \label{eq:example:concentration_consistency}
	\mathrm{x}_{A i} + \mathrm{x}_{B i} < 1,
\end{equation}
so that $\mathrm{x}_{C i}$, derived from \eqref{eq:example:concentrations}, is non-negative.
As illustrated in Section \ref{sec:towards:cost_function}, this condition can be accounted by penalizing the violation of \eqref{eq:example:concentration_consistency} in the loss function.
The following cost function is hence adopted
\begin{equation}
	L(\Phi_1, \Phi_2, \Phi_3) = \sum_{i =1}^{3} \Big[ 	\text{MSE}(\mathcal{I}_t; \Phi_i) + w \max( \mathrm{x}_{A i} + \mathrm{x}_{B i} - 1, 0) \Big],
\end{equation}
consisting of the sum of LSTM' MSE, see \eqref{eq:rnn:training_problem} and \eqref{eq:rnn:mse_def}, and of the physical consistency term weighted by $w$.
Here, we set $w = 0.05$.
Note that these LSTM are single-layer networks with $10$ units each, and they have been trained jointly by minimizing $L(\Phi_1, \Phi_2, \Phi_3)$  via the RMSProp optimizer.

The training set $\mathcal{I}_t$, the validation set $\mathcal{I}_v$, and the independent test set $\mathcal{I}_f$, are composed by $100$, $36$, and $1$ sequences, respectively, of length $T_s = 1000$ steps.
These trajectories have been collected with sampling time $\tau_s = 0.1\text{s}$ from a simulator of the chemical plant, excited with a multilevel pseudo-random signal.
The training procedure has been carried out with PyTorch for $1000$ epochs, discarding the first $T_w = 100$ steps of each sequence as washout period.

\begin{figure}[t]
\includegraphics[width=0.85 \linewidth]{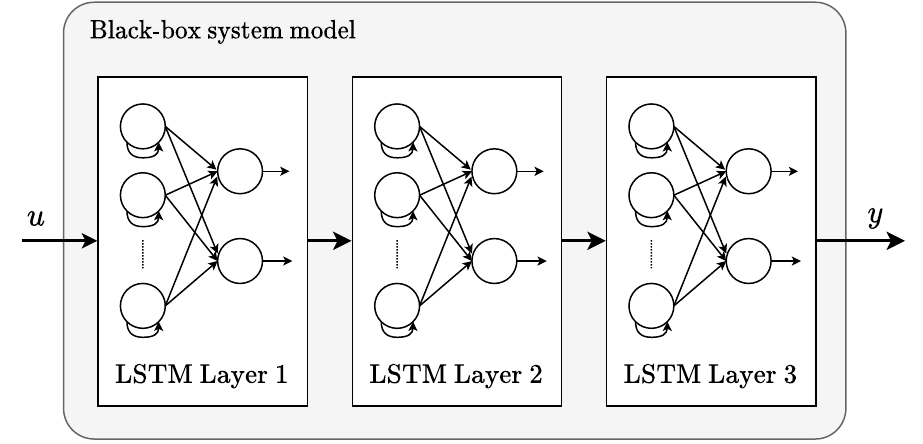}
\centering
\caption{Three-layer LSTM neural network without physical knowledge.}
\label{fig:nonphysics}
\end{figure}

To assess the advantages of the proposed physics-based mod\-el, the traditional black-box LSTM shown in Figure \ref{fig:nonphysics}, with $3$ layers of $10$ units each, has been used to perform black-box identification with the same data-set, training algorithm, and number of epochs.
Note that this non physics-based model has the same total number of units as the physics-based one.

\begin{figure}[t]
\includegraphics[width=0.475 \linewidth]{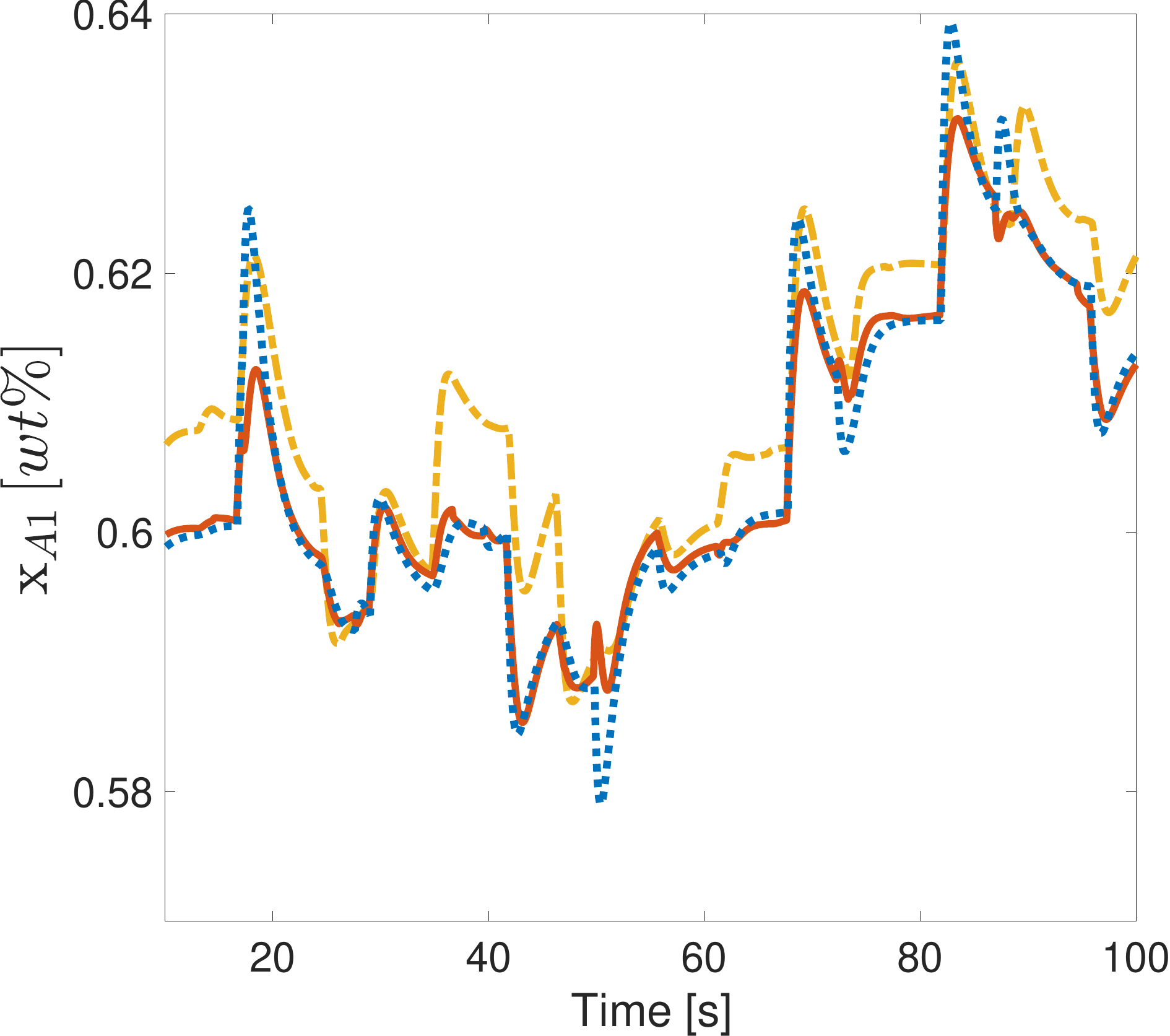} \hspace{2mm} 
\includegraphics[width=0.475 \linewidth]{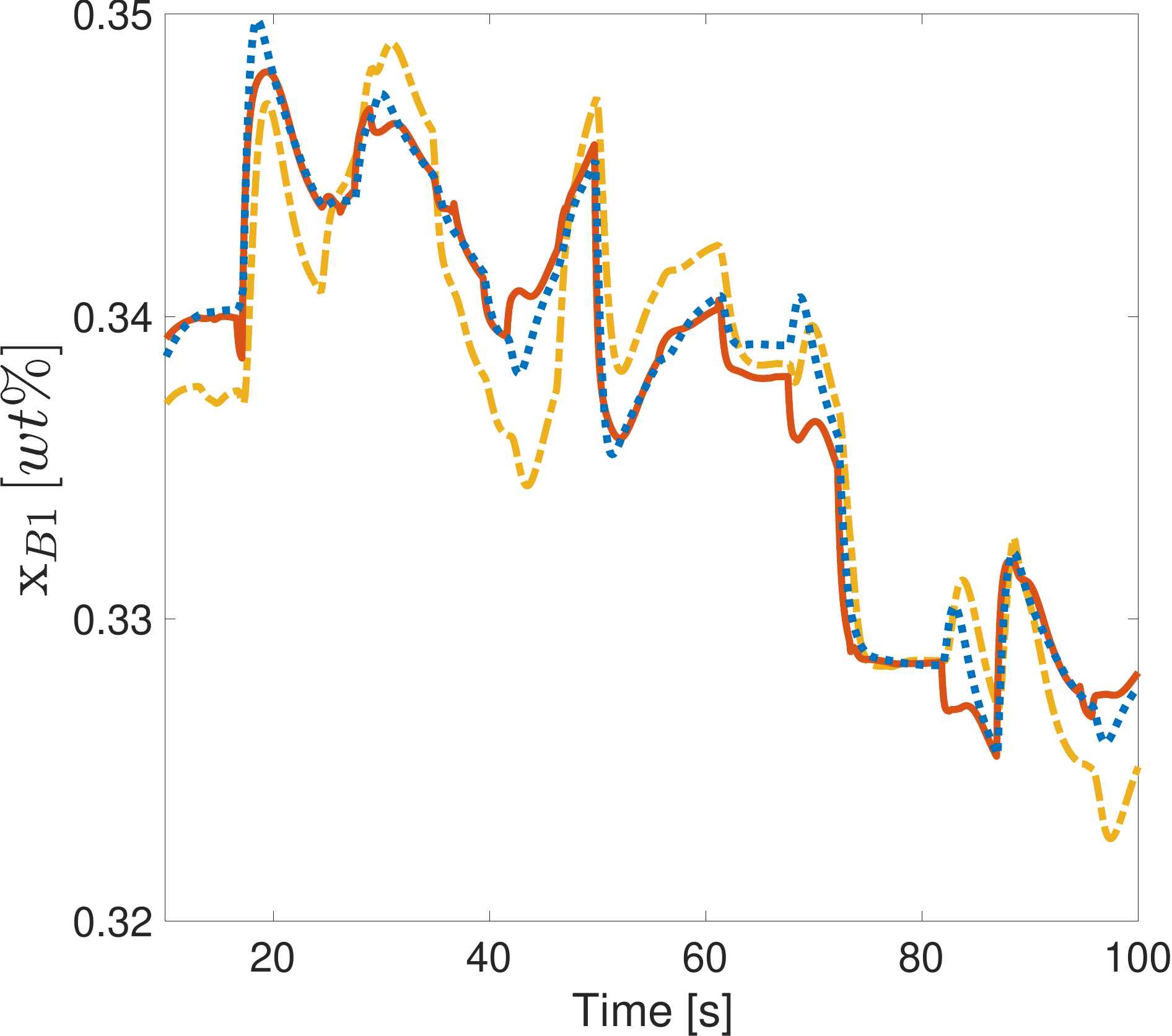} \vspace{2mm}\\
\includegraphics[width=0.475 \linewidth]{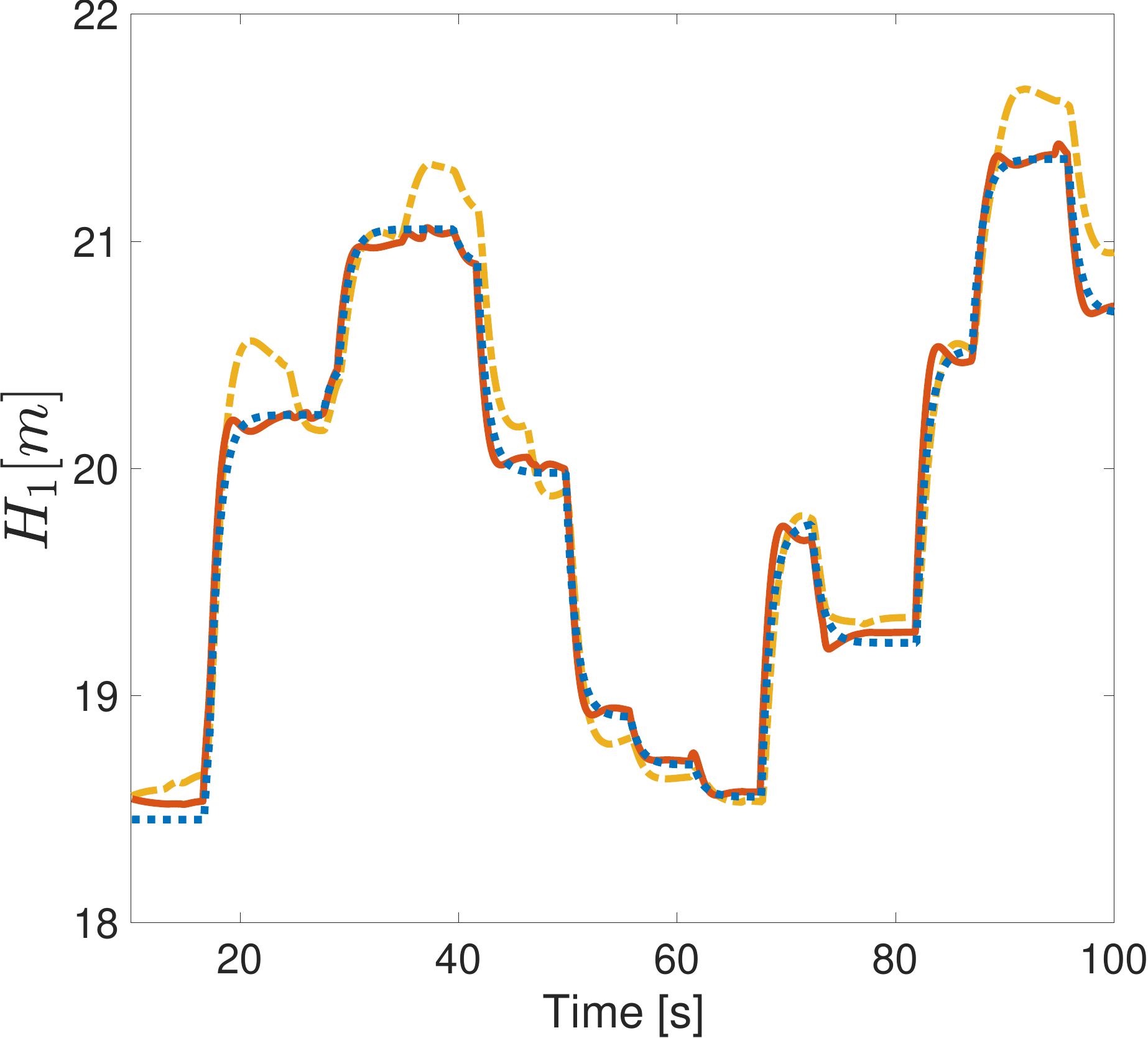}  \hspace{2mm} 
\includegraphics[width=0.475 \linewidth]{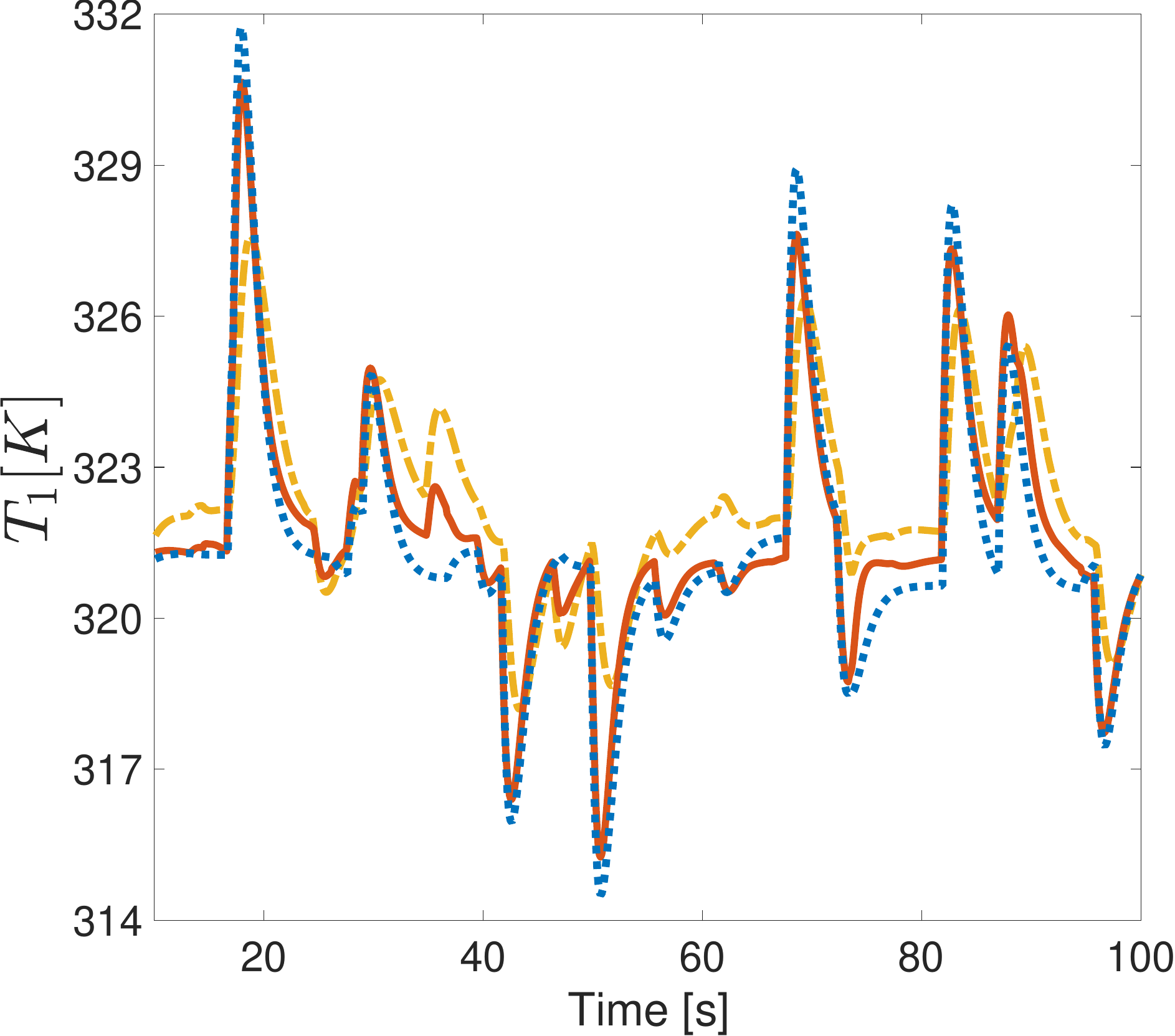}
\centering
\caption{Reactor $1$ modeling: physics-based LSTM (red solid line) and black-box LSTM (yellow dashed line) compared to the ground truth (blue dotted line) on the independent test sequence. 
Open-loop prediction of $\mathrm{x}_{A1}$ (top left), $\mathrm{x}_{B1}$ (top right), $H_1$ (bottom left), and $T_1$ (bottom right).}
\label{fig:modeling_performances}
\end{figure}

The modeling performances of the two networks on the independent test data-set are depicted in Figure \ref{fig:modeling_performances}.
For compactness, only the predictions of Reactor $1$'s outputs are illustrated; the overall performances of the two networks are reported later.
Figure \ref{fig:modeling_performances} qualitatively witnesses how the physics-based LSTM is, in general, more accurate than the black-box LSTM.

\begin{figure}[t]
	\centering
	\includegraphics[width=0.9\linewidth]{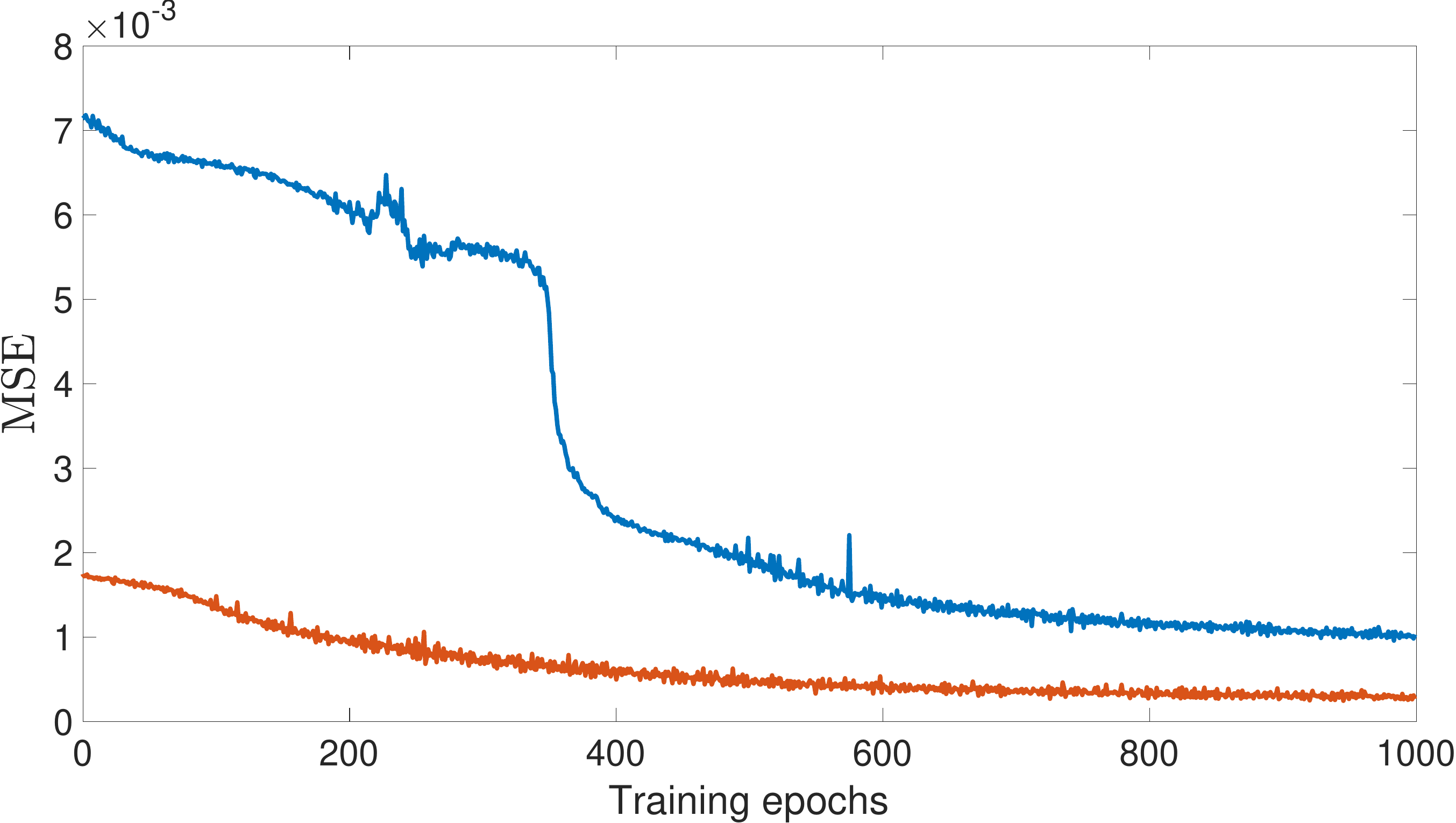}
	\caption{Evolution of the MSE during the training of the physics-based LSTM (red line) and of the black-box LSTM (blue line).}
	\label{fig:training_loss}
\end{figure}

To quantify the improvement, the FIT $[\%]$ index can be used. 
This measure is defined, for each of the $12$ outputs of the system, as
\begin{equation}
	\text{FIT} = 100 \bigg( 1 - \frac{1}{\lvert \mathcal{I}_f \lvert ( T_s - T_w )} \sum_{i \in \mathcal{I}_f} \sum_{k=T_w}^{Ts} \frac{ \| y_k^{\{i\}} - y_{m, k}^{\{i\}}  \|_{2}} {\| y_{m, k}^{\{i\}} - y_{avg} \|_{2}} \bigg),
\end{equation}
where $y_k^{\{i\}}$ denotes the model's open-loop prediction for the test sequence $i \in \mathcal{I}_f$, $y_{m, k}^{\{i\}}$  denotes the ground truth, and $y_{avg}$ the component-wise average value of the output $y_{m, k}^{\{i\}}$.
In Table \ref{tab:performances}, the FIT indexes achieved by the physics-based model are compared, for each output, to the indexes scored by the black-box model.
Notably, a significant performance improvement is achieved by the physics-based LSTM.
More specifically, the average inaccuracy of the physics model, defined as $1 - \text{FIT}$, is approximately half that of the black-box model.
Lastly, another advantage of the physics-based design is that it generally enjoys a smoother and faster convergence during training, as witnessed by Figure \ref{fig:training_loss}.

\begin{table}[t]
\centering
\begin{tabular}{ccc}
\toprule
Output & Black-box $[\%]$ &  Physics-based $[\%]$    \\ \midrule
$H_{1}$& 75.15  & 93.86  \\
$\mathrm{x}_{A1}$& 54.84 & 83.53  \\
$\mathrm{x}_{B1}$& 53.90  & 83.19  \\
$T_{1}$& 31.58  & 69.85   \\
$H_{2}$& 67.72  & 95.16   \\
$\mathrm{x}_{A2}$& 66.83  & 87.74  \\
$\mathrm{x}_{B2}$& 64.36  & 81.49  \\
$T_{2}$& 54.53  & 70.14   \\
$H_{3}$& 89.03  & 85.28   \\
$\mathrm{x}_{A3}$& 79.71  & 85.77  \\
$\mathrm{x}_{B3}$& 71.16  & 79.63  \\
$T_{3}$& 41.40  & 67.63   \\
\midrule
Overall & 62.52  & 82.67  \\
\bottomrule
\end{tabular}
\caption{Comparison of the FIT values achieved by the two models}
\label{tab:performances}
\end{table}

\section{Conclusions} \label{sec:conclusions}
In this paper, the use of Recurrent Neural Networks (RNN) for indirect data-driven control synthesis has discussed.
Despite in the last decades significant research efforts have been devoted to advocating the use of RNN for control, there are challenges and open theoretical questions that still need to be addressed.
From the perspective of the control system designer, the most relevant issues are guaranteeing the stability and robustness of the RNN model used to identify the plant, performing the safety verification of such model, and ensuring its interpretability and consistency with respect to the underlying physical laws characterizing the system.
Recent results and promising research directions towards these aims have been reported.

Many relevant issues, however, have not been discussed although, we believe, they well-deserve future attention from the control systems community.
Three of these main topics concern (\emph{i}) incremental training and adaptivity, which refer to the need to use the data, collected during the system's control operations, to improve the accuracy of the RNN model by means of additional training activities, see e.g. \cite{losing2018incremental}, while maintaining their stability properties; 
(\emph{ii}) stochastic control design based on probabilistic models learned from the data, see e.g. bayesian identification \cite{PigaBemporad}, Gaussian processes \cite{hewing2019cautious}, and probabilistic neural networks models \cite{hendriks2021deep}, extending the theoretical framework herein presented for deterministic models to these classes of probabilistic models; 
(\emph{iii}) deployment and testing of the proposed RNN-based control architecture to real systems, assessing its advantages compared to traditional control schemes.

\section*{Acknowledgements}
\hspace{0.8cm}

\begin{minipage}[c]{0.1\textwidth}
	\includegraphics[width=\textwidth]{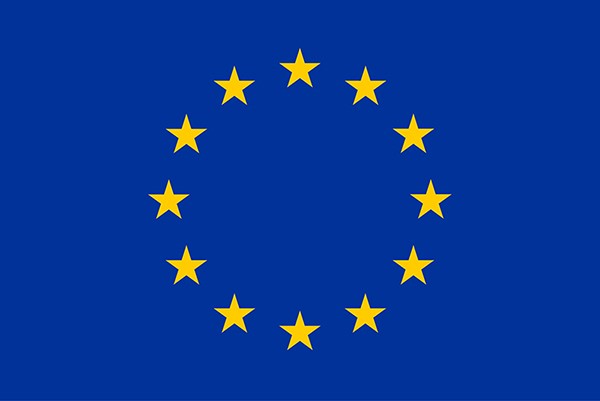}
	\label{fig:euflag}
\end{minipage}
\hspace{0.2cm}
\begin{minipage}[right]{0.3\textwidth}
	This project has received funding from the European Union’s Horizon 2020 research and innovation programme under the Marie Skłodowska-Curie grant agreement No. 953348
\end{minipage}
\vspace{2mm}

The authors are grateful to Dr. Alessio La Bella for his insightful advices.

\bibliographystyle{elsarticle-num}
\bibliography{referencesNN}

\appendix
\section{Sufficient conditions for RNN ISS and $\delta$ISS}
In this Appendix, the sufficient conditions for the ISS and $\delta$ISS of the RNN architectures discussed in Section \ref{sec:RNN} are briefly reported.
It is assumed that the input $u$ is unity-bounded, i.e. $\| u_k \|_\infty \leq 1$.
This is a quite general assumption when working with RNN, and can be easily fulfilled by means of a normalization procedure, see e.g. \cite{bonassi2021stability}.
\rev{In the following, being $A$ a matrix, $\| A \|_p$ is used to indicate its induced $p$-norm.}
Finally, let us remark that the following conditions are consistent with Proposition \ref{prop:stability}.

\subsection{NNARX \cite{bonassi2021nnarx}}
A sufficient condition for the is ISS and $\delta$ISS of the NNARX model described by \eqref{eq:rnn:nnarx_model} is that
\begin{equation} \label{eq:iss:condition}
	 \| U_0 \|_2 \cdot \| U_1 \|_2 - \frac{1}{ L_{\psi} \sqrt{N}} < 0.
\end{equation}

\subsection{ESN \cite{armenio2019model}}
A sufficient condition for the $\delta$ISS of ESN \eqref{eq:rnn:esn_statespace} is that the randomly-generated fixed weight matrix $U$ satisfies $\| U \|_2 < 1$ and
\begin{equation}
	\| U - W_y  U_o \|_2 - 1 < 0.
\end{equation}

\subsection{LSTM \cite{terzi2021learning}}
A sufficient condition for the ISS of the LSTM network \eqref{eq:rnn:lstm_statespace} is that 
\begin{equation} 
\bar{\sigma}_f + \bar{\sigma}_z \bar{\sigma}_i \| U_r \|_2 - 1 < 0,
 \end{equation}
where 
\begin{equation*} \label{eq:appendix:lstm:bounds}
\begin{aligned} 
	\bar{\sigma}_f &= \sigma \left( \| W_f \quad U_f \quad b_f \|_{\infty} \right), \\
	\bar{\sigma}_i &= \sigma \left( \| W_i \quad U_i \quad b_i \|_{\infty} \right), \\
	\bar{\sigma}_z &= \sigma \left(\| W_z \quad U_z \quad b_z \|_{\infty} \right), \\
	\bar{\phi}_r &= \phi \left(\| W_r \quad U_r \quad b_r \|_{\infty} \right).
\end{aligned}
\end{equation*}
Moreover, denoting by
\begin{equation*}
\begin{aligned}
	\alpha &= \frac{1}{4} \|U_f\|_2  \,  \frac{ \bar{\sigma}_i \bar{\phi}_r}{1-\bar{\sigma}_f}+   \bar{\sigma}_i   \|U_r\|_2   + \frac{1}{4} \|U_i\|_2  \,  \bar{\phi}_r, \\
	\bar{\phi}_h &= \phi \left( \frac{\bar{\sigma}_i \bar{\phi}_r}{1 - \bar{\sigma}_f} \right),
\end{aligned}
\end{equation*}
a sufficient conditions for the $\delta$ISS of the network is that the following pair of inequalities is satisfied
\begin{equation}
\begin{aligned}
	-1 + \bar{\sigma}_f + \alpha \bar{\sigma}_z + \frac{1}{4} \bar{\phi}_h \|U_z\|_2 
	- \frac{1}{4} \bar{\sigma}_f \bar{\phi}_h \| U_z \|_2 &< 0, \\
	\frac{1}{4} \bar{\sigma}_f \bar{\phi}_h \| U_z \|_2  - 1 &< 0.
\end{aligned}
\end{equation}

\subsection{GRU \cite{bonassi2021stability}}
A sufficient condition for the $ISS$ of the GRU network \eqref{eq:rnn:gru_statespace} is that
\begin{equation} \label{eq:single:iss:condition}
  \| U_r \|_\infty \, \bar{\sigma}_f - 1 < 0,
\end{equation}
where
\begin{equation*}
  \bar{\sigma}_f = \sigma \left( \| W_f \quad U_f \quad b_f \|_\infty \right).
\end{equation*}
Moreover, a sufficient condition for the $\delta$ISS is that
\begin{equation}
 \| U_r \|_\infty \left( \frac{1}{4} \| U_f \|_\infty + \bar{\sigma}_f \right) + \frac{1}{4} \frac{1 + \bar{\phi}_r}{1 - \bar{\sigma 
 }_z} \| U_z \|_\infty - 1 < 0,
\end{equation}
where
\begin{equation*}
\begin{aligned}
	\bar{\sigma}_z &= \sigma( \| W_z \quad U_z \quad b_z \|_\infty ) < 1, \\
	\bar{\phi}_r &= \phi( \| W_r \quad U_r \quad b_r \|_\infty ) < 1.
\end{aligned}
\end{equation*}

\end{document}